\documentclass[twocolumn,
              superscriptaddress,
              prd, aps,
              showkeys,
              showpacs,
              nofootinbib,
              notitlepage,
              floatfix,
              longbibliography,
              ]{revtex4-2}

\usepackage[english]{babel}
\usepackage[letterpaper,top=2cm,bottom=2cm,left=2cm,right=2cm,marginparwidth=1.75cm]{geometry}
\usepackage{amsmath}
\usepackage{graphicx}
\usepackage{booktabs}
\usepackage{siunitx}
\usepackage{upgreek}
\usepackage{xspace}
\usepackage[colorlinks=true,citecolor=blue,filecolor=blue,linkcolor=blue,urlcolor=blue,pdftex]{hyperref}
\usepackage[dvipsnames]{xcolor}
\usepackage{lipsum}
\usepackage{multirow}
\usepackage{hyperref}
\usepackage[capitalise]{cleveref}
\crefname{section}{Sec.}{Secs.}
\crefname{table}{Tab.}{Tabs.} 
\usepackage{subfigure}
\usepackage{adjustbox}
\usepackage{textcomp}
\usepackage{orcidlink} 
\usepackage{currfile} 

\DeclareSIUnit\eVperc{\eV\per\clight}
\DeclareSIUnit\clight{\text{\ensuremath{c}}}

\begin{document}

\title{XENONnT WIMP Search: Signal \& Background Modeling and Statistical Inference}

\newcommand*{\comment}{\textcolor{red}}
\newcommand*{\needs}{\textcolor{red}}
\newcommand{\rntwotwozero}{\ensuremath{^{220}\mathrm{Rn}}\xspace}
\newcommand{\toymc}{toy-MC\xspace}
\newcommand{\toymcs}{toy-MCs\xspace}
\newcommand{\cevns}{CE\ensuremath{\nu}NS\xspace}
\newcommand{\x}{\ensuremath{\mathrm{X}}\xspace}
\newcommand{\y}{\ensuremath{\mathrm{Y}}\xspace}
\newcommand{\z}{\ensuremath{\mathrm{Z}}\xspace}
\newcommand{\R}{\ensuremath{\mathrm{R}}\xspace}
\newcommand{\radius}{\ensuremath{\mathrm{R}}\xspace}
\newcommand{\wimp}{\ensuremath{\mathrm{WIMP}}\xspace}

\newcommand{\expectationACnonwirenominaloffall}{ $2.0 \pm 0.6 $ }
\newcommand{\expectationACwirenominalonall}{ $2.3 \pm 0.7 $ }
\newcommand{\expectationatnunominalbothall}{ $0.05 \pm 0.02 $ }
\newcommand{\expectationcevnsnominalbothall}{ $0.19 \pm 0.06 $ }
\newcommand{\expectationernominalbothall}{ $134.5 $ }
\newcommand{\expectationradiogenicnominalbothall}{ $0.8 \pm 0.4 $ }
\newcommand{\expectationradiogenicXnominalbothall}{ $0.31 \pm 0.16 $ }
\newcommand{\expectationwallnominalbothall}{ $14 \pm 3 $ }
\newcommand{\expectationercalibrationnominalboth}{ $2062 \pm 210 $ }
\newcommand{\expectationparonenominal}{ $0.0 $ }
\newcommand{\expectationpartwonominal}{ $0.0 $ }
\newcommand{\expectationACnonwirebestoffall}{ $2.1 \pm 0.6 $ }
\newcommand{\expectationACwirebestonall}{ $2.3 \pm 0.6 $ }
\newcommand{\expectationatnubestbothall}{ $0.04 \pm 0.02 $ }
\newcommand{\expectationcevnsbestbothall}{ $0.19 \pm 0.06 $ }
\newcommand{\expectationerbestbothall}{ $135_{ -11  }^{ +12 }$ }
\newcommand{\expectationradiogenicbestbothall}{ $0.8 \pm 0.4 $ }
\newcommand{\expectationradiogenicXbestbothall}{ $0.30 \pm 0.15 $ }
\newcommand{\expectationwallbestbothall}{ $12 \pm 2 $ }
\newcommand{\expectationercalibrationbestboth}{ $2052 \pm 44 $ }
\newcommand{\expectationparonebest}{ $0.4 \pm 0.2 $ }
\newcommand{\expectationpartwobest}{ $-1.8_{ -0.6  }^{ +0.7 }$ }

\makeatletter
\newcommand{\fmarki}{*}
\newcommand{\fmarkii}{\ensuremath{\dagger}}
\newcommand{\fmarkiii}{\ensuremath{\ddagger}}
\newcommand{\fmarkiv}{\ensuremath{\mathsection}}
\newcommand{\fmarkv}{\ensuremath{\mathparagraph}}
\newcommand{\fmarkvi}{\ensuremath{\|}}
\newcommand{\fmarkvii}{**}
\newcommand{\fmarkviii}{\ensuremath{\dagger\dagger}}
\newcommand{\fmarkix}{\ensuremath{\ddagger\ddagger}}

\def\@fnsymbol#1{{\ifcase#1\or \fmarki\or \fmarkii\or \fmarkiii\or \fmarkiv\or \fmarkv\or \fmarkvi\or \fmarkvii\or \fmarkviii\or \fmarkix \else\@ctrerr\fi}}
\makeatother
\newcommand{\bologna}{\affiliation{Department of Physics and Astronomy, University of Bologna and INFN-Bologna, 40126 Bologna, Italy}}
\newcommand{\chicago}{\affiliation{Department of Physics \& Kavli Institute for Cosmological Physics, University of Chicago, Chicago, IL 60637, USA}}
\newcommand{\coimbra}{\affiliation{LIBPhys, Department of Physics, University of Coimbra, 3004-516 Coimbra, Portugal}}
\newcommand{\columbia}{\affiliation{Physics Department, Columbia University, New York, NY 10027, USA}}
\newcommand{\lngs}{\affiliation{INFN-Laboratori Nazionali del Gran Sasso and Gran Sasso Science Institute, 67100 L'Aquila, Italy}}
\newcommand{\mainz}{\affiliation{Institut f\"ur Physik \& Exzellenzcluster PRISMA$^{+}$, Johannes Gutenberg-Universit\"at Mainz, 55099 Mainz, Germany}}
\newcommand{\mpik}{\affiliation{Max-Planck-Institut f\"ur Kernphysik, 69117 Heidelberg, Germany}}
\newcommand{\munster}{\affiliation{Institut f\"ur Kernphysik, Westf\"alische Wilhelms-Universit\"at M\"unster, 48149 M\"unster, Germany}}
\newcommand{\nikhef}{\affiliation{Nikhef and the University of Amsterdam, Science Park, 1098XG Amsterdam, Netherlands}}
\newcommand{\nyuad}{\affiliation{New York University Abu Dhabi - Center for Astro, Particle and Planetary Physics, Abu Dhabi, United Arab Emirates}}
\newcommand{\purdue}{\affiliation{Department of Physics and Astronomy, Purdue University, West Lafayette, IN 47907, USA}}
\newcommand{\rice}{\affiliation{Department of Physics and Astronomy, Rice University, Houston, TX 77005, USA}}
\newcommand{\stockholm}{\affiliation{Oskar Klein Centre, Department of Physics, Stockholm University, AlbaNova, Stockholm SE-10691, Sweden}}
\newcommand{\subatech}{\affiliation{SUBATECH, IMT Atlantique, CNRS/IN2P3, Universit\'e de Nantes, Nantes 44307, France}}
\newcommand{\torino}{\affiliation{INAF-Astrophysical Observatory of Torino, Department of Physics, University  of  Torino and  INFN-Torino,  10125  Torino,  Italy}}
\newcommand{\ucsd}{\affiliation{Department of Physics, University of California San Diego, La Jolla, CA 92093, USA}}
\newcommand{\wis}{\affiliation{Department of Particle Physics and Astrophysics, Weizmann Institute of Science, Rehovot 7610001, Israel}}
\newcommand{\zurich}{\affiliation{Physik-Institut, University of Z\"urich, 8057  Z\"urich, Switzerland}}
\newcommand{\paris}{\affiliation{LPNHE, Sorbonne Universit\'{e}, CNRS/IN2P3, 75005 Paris, France}}
\newcommand{\freiburg}{\affiliation{Physikalisches Institut, Universit\"at Freiburg, 79104 Freiburg, Germany}}
\newcommand{\napels}{\affiliation{Department of Physics ``Ettore Pancini'', University of Napoli and INFN-Napoli, 80126 Napoli, Italy}}
\newcommand{\nagoya}{\affiliation{Kobayashi-Maskawa Institute for the Origin of Particles and the Universe, and Institute for Space-Earth Environmental Research, Nagoya University, Furo-cho, Chikusa-ku, Nagoya, Aichi 464-8602, Japan}}
\newcommand{\laquila}{\affiliation{Department of Physics and Chemistry, University of L'Aquila, 67100 L'Aquila, Italy}}
\newcommand{\tokyo}{\affiliation{Kamioka Observatory, Institute for Cosmic Ray Research, and Kavli Institute for the Physics and Mathematics of the Universe (WPI), University of Tokyo, Higashi-Mozumi, Kamioka, Hida, Gifu 506-1205, Japan}}
\newcommand{\kobe}{\affiliation{Department of Physics, Kobe University, Kobe, Hyogo 657-8501, Japan}}
\newcommand{\kit}{\affiliation{Institute for Astroparticle Physics, Karlsruhe Institute of Technology, 76021 Karlsruhe, Germany}}
\newcommand{\tsinghua}{\affiliation{Department of Physics \& Center for High Energy Physics, Tsinghua University, Beijing 100084, P.R. China}}
\newcommand{\ferrara}{\affiliation{INFN-Ferrara and Dip. di Fisica e Scienze della Terra, Universit\`a di Ferrara, 44122 Ferrara, Italy}}
\newcommand{\groningen}{\affiliation{Nikhef and the University of Groningen, Van Swinderen Institute, 9747AG Groningen, Netherlands}}
\newcommand{\westlake}{\affiliation{Department of Physics, School of Science, Westlake University, Hangzhou 310030, P.R. China}}
\newcommand{\shenzhen}{\affiliation{School of Science and Engineering, The Chinese University of Hong Kong, Shenzhen, Guangdong, 518172, P.R. China}}
\newcommand{\coimbrapoli}{\affiliation{Coimbra Polytechnic - ISEC, 3030-199 Coimbra, Portugal}}
\newcommand{\uniheidelberg}{\affiliation{Physikalisches Institut, Universit\"at Heidelberg, Heidelberg, Germany}}
\newcommand{\roma}{\affiliation{INFN-Roma Tre, 00146 Roma, Italy}}
\newcommand{\bucknell}{\affiliation{Department of Physics \& Astronomy, Bucknell University, Lewisburg, PA, USA}}


\author{E.~Aprile\,\orcidlink{0000-0001-6595-7098}}\columbia
\author{J.~Aalbers\,\orcidlink{0000-0003-0030-0030}}\groningen
\author{K.~Abe\,\orcidlink{0009-0000-9620-788X}}\tokyo
\author{S.~Ahmed Maouloud\,\orcidlink{0000-0002-0844-4576}}\paris
\author{L.~Althueser\,\orcidlink{0000-0002-5468-4298}}\munster
\author{B.~Andrieu\,\orcidlink{0009-0002-6485-4163}}\paris
\author{E.~Angelino\,\orcidlink{0000-0002-6695-4355}}\torino\lngs
\author{D.~Ant\'on~Martin\,\orcidlink{0000-0001-7725-5552}}\chicago
\author{F.~Arneodo\,\orcidlink{0000-0002-1061-0510}}\nyuad
\author{L.~Baudis\,\orcidlink{0000-0003-4710-1768}}\zurich
% ...
\collaboration{XENON Collaboration}\email[]{xenon@lngs.infn.it}\noaffiliation

%

\date{\today}

\begin{abstract}
    The XENONnT experiment searches for weakly-interacting massive particle (WIMP) dark matter scattering off a xenon nucleus.
    In particular, XENONnT uses a dual-phase time projection chamber with a 5.9-tonne liquid xenon target, detecting both scintillation and ionization signals to reconstruct the energy, position, and type of recoil.
    A blind search for nuclear recoil WIMPs with an exposure of 1.1 tonne-years ($4.18\;\mathrm{t}$ fiducial mass) yielded no signal excess over background expectations, from which competitive exclusion limits were derived on WIMP-nucleon elastic scatter cross sections, for WIMP masses ranging from 6\,GeV/$c^2$ up to the TeV/$c^2$ scale.
    This work details the modeling and statistical methods employed in this search.
    By means of calibration data, we model the detector response, which is then used to derive background and signal models.
    The construction and validation of these models is discussed, alongside additional purely data-driven backgrounds.
    We also describe the statistical inference framework, including the definition of the likelihood function and the construction of confidence intervals.
    
\end{abstract}


\maketitle

\section{Introduction}
\label{intro}

Astrophysical and cosmological observations at various scales provide clear evidence for the existence of dark matter (DM) as a fundamental building block of our universe \cite{Bertone_dm_history_2018, Planck_2018_6}.
Numerous extensions of the standard model of particle physics predict additional fundamental particles as potential candidates for DM \cite{Bertone_2005_dm_candidates, Roszkowski2018_dm_candidates}.
Measuring the direct interaction of particles from the DM halo of the Milky Way with a suitable detector could provide conclusive evidence for the particle DM hypothesis.
The XENONnT experiment aims to detect a signal from weakly interacting massive particles (WIMPs) elastically scattering off xenon nuclei with a detector operated underground at the Istituto Nazionale di Fisica Nucleare (INFN) Laboratori Nazionali del Gran Sasso (LNGS) in Italy.
The collaboration published the first WIMP search results from a data-taking period between July $6^\mathrm{th}$ and November $10^\mathrm{th}$ 2021 with a total live time of $95.1$ days and a fiducial mass of $(4.18\pm0.13)\;\mathrm{t}$, referred to as science run 0 (SR0).
We performed a blind analysis and observed no significant excess above background expectations  \cite{aprile_search_2023}.
This led to a minimum upper limit on the spin-independent WIMP-nucleon cross section of $2.58 \times 10^{-47}\;\mathrm{cm}^2$ for a WIMP mass of \SI[per-mode=symbol]{28}{\giga\eVperc\squared} at $90\%$ conﬁdence level (CL).

The detector used for the XENONnT experiment is a dual-phase xenon time projection chamber (TPC).
A particle interaction in the active \SI{5.9}{t} liquid xenon (LXe) target results in prompt vacuum ultraviolet scintillation light as well as free ionization electrons.
The instantaneous light signal, called the S1 signal, is detected by two arrays of photomultiplier tubes (PMTs) at the top and bottom of the approximately cylindrical TPC.
An electric field (\SI{23}{V/cm} average electric drift field during SR0) is applied to the target volume to drift the ionization electrons toward the liquid xenon surface.
Here, they are extracted and accelerated into a xenon gas volume above the liquid via a stronger electric field (\SI{2.9}{kV/cm} electric extraction field in liquid during SR0).
The kinetic energy of the accelerated electrons in the gas phase is sufficient for the emission of electroluminescent light, which is proportional to the number of extracted electrons.
This second light signal is referred to as the S2 signal.
The distribution of the S2 signal detected by the PMTs in the top array is used to infer the position of the initial interaction in the horizontal plane $\mathrm{(\x, \y)}$ parallel to the liquid surface, which defines the radial coordinate $\mathrm{\radius=\sqrt{\x^2+\y^2}}$.
The vertical component \z follows from the drift time of electrons, which is determined as the time difference between the S1 and S2 signals.
This three-dimensional position reconstruction of the interaction vertex enables the discrimination of multi-site events as well as the selection of an inner fiducial volume (FV) within the TPC.
This volume benefits from a particularly low background level thanks to the excellent self-shielding properties of LXe, given its high density.
Moreover, the relative magnitudes of the S1 and S2 signals can be used to determine if a particle interacted with the xenon nucleus (expected from WIMPs) or xenon shell electrons (typical of the dominant sources of background), i.e. nuclear recoil (NR) or electronic recoil (ER) events.
The electric drift and extraction fields are established by three parallel-wire electrodes (cathode, gate, and anode from bottom to top).
To reduce wire sagging, two and four horizontal \textit{perpendicular wires} support the gate and anode electrodes, respectively.

The TPC is nested in an active neutron veto, which in turn is housed in an active muon veto. Both veto systems are water Cherenkov detectors.
More details on the TPC, the veto systems, and other subsystems of the detector are provided in \cite{xenonnt_2024_intrumentation, aprile_field_cage_2023, xe_neutron_veto}.

The data analysis chain of the WIMP DM search in XENONnT is subdivided into two major parts, like in XENON1T \cite{xenon1t_2019_ap1,xenon1t_2019_ap2}.
The first part is discussed in~\cite{nT_sr0_analysispaper_I} and comprises signal and event reconstruction, corrections, event selection, and energy scale calibration.
Here we present the second part of the analysis chain:
The detector response to the different interaction types is modeled based on calibration data and discussed in \cref{sec:detector_response}.
The derived best-fit models are important inputs for the definition of signal and background models, detailed in \cref{sec:signal_and_background}.
This section also covers data-driven background models that do not rely on the detector response.
Finally, the statistical methods used to compute constraints on DM given the XENONnT data are discussed in \cref{sec:inference}.

\section{Modeling the detector response}
\label{sec:detector_response}
Modeling the detector response to energy depositions allows for a powerful discrimination between NR signal and ER background in S1--S2 space.
NR events are produced by particles scattering elastically off xenon nuclei, while in ER events, particles scatter off xenon shell electrons.
The different energy loss processes involved after the two types of recoil cause ER and NR events to form separate populations in S1--S2 space, which is exploited in the WIMP DM search.
The modeling is performed in the region of interest (ROI) designated for the SR0 WIMP search, defined by $\mathrm{cS1}$ in [0, 100]\,PE (PE = photoelectron) and $\mathrm{cS2}$ in [$10^{2.1}$, $10^{4.1}$]\,PE.
This ROI is selected to contain the majority of our expected signal (c.f. \cref{fig:template_contours,fig:wimp_template_fig}).
We build the models using calibration data before unblinding the WIMP search dataset.

In this section, we describe the TPC response model to energy depositions in LXe via ER or NR interactions, which is obtained from fits to calibration data.
This detector response model is separated into two parts.
The first one is the empirically parametrized LXe emission model, which describes the production processes of the detectable quanta, i.e. the conversion of deposited energy into scintillation photons and ionization electrons.
The second part is the detector reconstruction model, which covers the conversion from the produced photons and electrons into the observed and spacetime-dependence corrected S1 and S2 signals (cS1 and cS2).
Due to its complexity, it is unfeasible in the fit to simulate processes of photon and electron propagation on a per-quantum basis.
Instead, toy-Monte-Carlo (\toymc) simulations of the detector reconstruction model are used which sample from effective maps provided either by data-driven analyses described in~\cite{nT_sr0_analysispaper_I} or by simulation-driven analyses using the XENONnT Monte Carlo (MC) framework \cite{ramirez_thesis, zhu_thesis}.
All model parameters in the simulations are fit to both ER and NR calibration data.
The two parts of the detector response model together with the \toymc simulation workflow are described in \cref{sec:microphys,sec:recon}, and the fit to ER and NR calibration data is detailed in \cref{sec:fit}.

\subsection{Liquid Xenon Emission Model}
\label{sec:microphys}

The production of quanta from energy depositions in a xenon target is complex and lacks a comprehensive description derived from first principles.
Thus, a semi-empirical parametrization of the emission model is commonly used.
The parametrization used in XENONnT SR0 is similar to XENON1T~\cite{xenon1t_2019_ap2}, which is based on the Noble Element Simulation Technique (NEST) models described in~\cite{szydagis2011nest, Lenardo_2015}.
The simulation workflow of the emission model is described in detail in \cref{sec:appendix_er_emission_model,sec:appendix_nr_emission_model}, and is summarized in the following.

In an ER event, the recoil energy is transmitted into the excitation and ionization of xenon atoms.
Recoiling xenon nuclei from an NR event, on the other hand, lose a significant amount (roughly 80\,\%) of their kinetic energy to atomic motion in collisions with other xenon atoms \cite{szydagis2011nest}. This thermalization process is undetectable in LXe TPCs, resulting in an effective signal loss. For both recoil types, the total number of detectable quanta $N_{\mathrm{q}}$ equals the sum of the number of produced excitons $N_{\mathrm{ex}}$ and ions $N_{\mathrm{i}}$.
A fraction of ions and electrons recombine, resulting in the production of additional excitons. An exciton can combine with a ground-state xenon atom forming an excimer, which de-excites, producing a scintillation photon.
In an event, $N_\gamma$ photons are generated, part of which are then detected by PMTs as the $\mathrm{S1}$ signal. The $N_{\mathrm{e}}$ electrons that do not recombine are drifted to the liquid-gas interface and eventually form an $\mathrm{S2}$ signal.

In the SR0 WIMP search ROI, 
ER interactions have a much smaller ratio of produced excitons to electron-ion pairs than NR interactions (by a factor of up to $\sim$10 in the ROI), while the fraction of electron-ion pairs that recombine is similar for both interaction types given the same number of detectable quanta.
The difference in exciton-to-ion ratio dominates, leading to larger ratios of S2 to S1 areas for ER interactions, which is the principle behind ER-NR discrimination in LXe TPCs.
We parametrize the emission model for ER interactions as in \cite{xenon1t_2019_ap2}, with a constant exciton-to-ion ratio, and a mean recombination fraction following the Thomas-Imel box model~\cite{Thomas:1987zz} with an additional Fermi-Dirac term.
For NR interactions, the emission model follows the parametrization used in the NEST v1 model detailed in \cite{Lenardo_2015}.
While there are newer versions of fits available in the NEST framework that use a different parametrization for NR interactions (which we refer to as NEST v2 in the following), we choose to stay with the previous version (referred to as NEST v1).
The model is simpler, fits our data well, and the best-fit values and uncertainties of parameters of NEST v2 were not published at the time of the analysis.

\subsection{Detector Reconstruction Model}
\label{sec:recon}
The detector reconstruction model covers all aspects of the signal reconstruction, starting from the produced scintillation photons and ionization electrons up to the measured S1 and S2 signal sizes.
The XENONnT detector reconstruction model is almost identical to the one presented in \cite{xenon1t_2019_ap2} and is briefly outlined here.
The detailed simulation workflow of the detector reconstruction model is described in \cref{sec:appendix_detector_response_model}.

For S1 signals, the spatial dependence of geometric effects during photon propagation, the photon detection efficiency of the PMTs, and the effect of double photoelectron emission (DPE) from the PMT photocathode~\cite{LOPEZPAREDES201856, Faham_2015} are all modeled in the simulations.
For S2 signals, we model the loss of electrons along their drift path due to attachment to electronegative impurities as well as due to electric field effects close to the TPC wall, the efficiency of electron extraction from the liquid into the gas phase, and the single-electron gain of extracted electrons to detected PE via the electroluminescence process in gas.
In addition to these physical processes, we also account for the influence of software reconstruction effects on the signals.
For both signal types, the software reconstruction process can introduce biases and fluctuations in the recorded signal size.
We also account for the S1 signal reconstruction efficiency, which vanishes at about 3 PE, as we require that at least 3 PMTs detect a photon within $\pm$50\,ns around the maximal amplitude.
Finally, we model the uncertainty of the event position reconstruction and the acceptances of data quality selections \cite{nT_sr0_analysispaper_I}.

The modeling of ER interactions relies on calibration data from sources dissolved in xenon, in particular $^{220}$Rn and $^{37}$Ar.
The $^{220}$Rn progeny $^{212}$Pb undergoes a $\beta$-decay with a Q-value at about 0.6\,MeV~\cite{Jorg_2023}, giving an approximately flat ER energy spectrum in the WIMP ROI, i.e. below about $10$\,keV.
This dataset is used for the fit of the ER model.
The $^{37}$Ar source provides a mono-energetic ER peak when its K-shell electron capture process leads to a fast cascade of X-rays and Auger-Meitner electrons, with a total energy of about 2.8\,keV \cite{aprile2023low}.
Both ER sources provide single-scatter (SS) events, producing one S1 and one S2 signal for the total deposited energy, and the above-described detector reconstruction processes are simulated accordingly.

The NR model is calibrated using an external $^{241}$AmBe neutron source (referred to as AmBe source in the following).
It emits fast neutrons via the $^{9}$Be$(\alpha, n)^{12}$C capture reaction.
Neutrons often scatter multiple times in the LXe target in so-called multi-scatter (MS) events, creating distinct energy depositions at each interaction site, thus the topologies of neutron events must be considered in the modeling.
Since the neutrons travel at about 10\,\% of the speed of light and the S1 width is $\mathcal{O}(\mathrm{100\,ns})$, the S1 signals from the separate energy depositions in a MS event get reconstructed into one merged S1 signal.
The S2 signals might be distinguished based on the separation in spacetime between the energy depositions of the individual scatters, but can get (partially) merged if the interaction sites are close together.
Due to the low drift field of $23\,\mathrm{V/cm}$ in XENONnT SR0, the \z separation resolution between two S2 signals is reduced with respect to the design because of the increased drift time $\mathcal{O}(\mathrm{ms})$ and consequently larger S2 spread $\mathcal{O}(\mathrm{\upmu s})$ due to the diffusion of electrons.
This effect is accounted for in the fit of the NR model to the neutron calibration data: we first produce the photons and electrons for each energy deposition in a MS event separately, based on the emission model.
While the primary scintillation photons are summed to produce the merged S1 signal, we only sum S2 signals that will become part of the largest S2 signal after going through the remaining detector reconstruction processes.
This information is provided by PMT waveform simulations of neutron scatters from the AmBe source, using the Monte Carlo (MC) framework of XENONnT~\cite{ramirez_thesis, zhu_thesis} (see also \cref{subsec:neutron_bkg} for details on neutron simulations).
The results of these waveform simulations are used as inputs to the \toymc simulations and fitting framework.
Data selections that remove MS events (both resolved and unresolved) are directly applied to these inputs, because their acceptances can only be determined from full simulations, and are correspondingly dropped from the selection acceptance curves for the NR fit.

\begin{figure}[t!]
    \centering
    \includegraphics[width=\columnwidth]{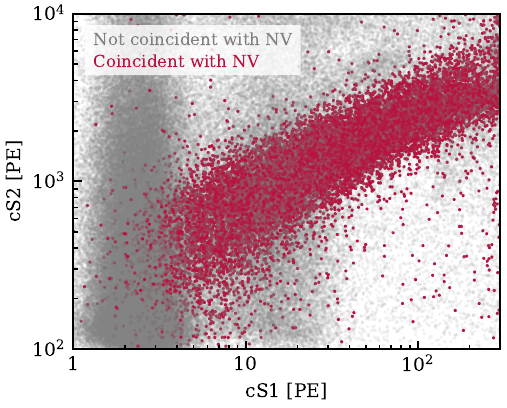}
    \caption{AmBe neutron calibration events with (red) and without (gray) a coincident signal in the neutron veto (NV). Selecting coincident events ensures a clean nuclear recoil sample for detector response modeling. The accidental coincidence population (cS1 below 5\,PE) and misidentified single-electron S1s (cS1 between 10 and 30\,PE) are visible in the gray population. After applying additional data quality selections, approximately 2000 events (\cref{fig:ernrgofs}, right) are used for band fitting.}
    \label{fig:ambe_selection}
\end{figure}

\subsection{Fit to Calibration Data}
\label{sec:fit}

\begin{figure*}[ht!]
    \centering
        \includegraphics[width=\textwidth]{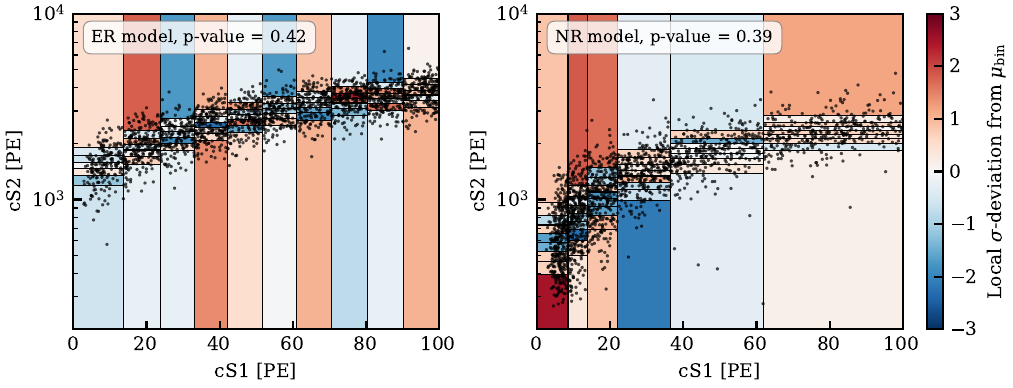}
    \caption{Comparison between calibration data and the best-fit ER (left) and NR (right) models. The equiprobable binning for the 2D binned Poisson likelihood $\chi^2$ goodness-of-fit tests is shown. The color scale indicates the deviation of the number of data points (overlaid as black dots) in each bin from the best-fit model expectation $\mu_\mathrm{bin}$ in units of sigma, which is 29.3 for the ER model and 36.9 for the NR model.}
    \label{fig:ernrgofs}
\end{figure*}

The parameters in the full detector response model are fitted to $^{220}$Rn, $^{37}$Ar, and AmBe calibration data.
The ROI selection is the same as the SR0 WIMP search region defined in \cref{sec:microphys}.
All data quality selections are applied to the calibration data within the ROI.
For the AmBe neutron data, an additional selection is applied using coincident gammas detected in the neutron veto.
In $^{9}$Be$(\alpha, n)^{12}$C capture reactions in the AmBe source, $^{12}$C is left in the first excited state after neutron emission in about 50\,\% of the cases, estimated with neutron veto data~\cite{danielwphd}.
The $^{12}$C excited state promptly decays via the emission of a 4.44\,MeV gamma ray, coincident with the neutron emission.
The requirement of the time coincidence within $250\,\mathrm{\upmu s}$ between the 4.44\,MeV gamma ray detected in the neutron veto and the event in the TPC, leads to a clean calibration dataset, with only about one expected event from accidental coincidences between the two detectors in the dataset ($\sim$0.05\,\% of the resulting sample). The events recorded in the ROI that pass (fail) the neutron veto coincidence selection are shown in red (gray) before applying any additional data quality selections in~\cref{fig:ambe_selection}.

For the fit to the ER data, an additional background component is added to account for accidental coincidence (AC, see \autoref{subsec:ac_bkg} for details) events formed by a wrong pairing of S1 and S2 signals in the calibration dataset.

After all selections, the ${}^{220}$Rn and AmBe datasets consist of $\sim$2000 events each.
We use a downsampled dataset of 10\,000 $^{37}$Ar events to perform a combined ${}^{220}$Rn-${}^{37}$Ar ER fit.
With the downsampling we avoid overconstraining the fit to the narrow, very low energy range of $^{37}$Ar.


For the fit, an affine-invariant Markov chain Monte Carlo (MCMC) algorithm \cite{emcee,emceev3} is used to sample from the high-dimensional posterior distribution of the parameters.
In each step of the sampling, a GPU-accelerated \toymc simulation is performed for every set of the ER and NR model parameters, following the steps described in the previous two sections, producing a model of the detector response in the space of corrected signals, i.e. $\mathrm{cS1}$--$\mathrm{cS2}$.
The model parameters are described in detail in the appendix and are listed in \cref{tab:bbf_param_summary}.
These \toymc models are then compared to calibration data by calculating the binned Poisson likelihood in $\mathrm{cS1}$--$\mathrm{cS2}$ space.
The likelihoods are multiplied by the prior distributions of the parameters to yield the posteriors.
After sufficient iterations of the MCMC, the samples in the chain then converge to the posterior distribution of the parameters.

\begin{figure*}[t]
    \centering
    \includegraphics[width=\textwidth]{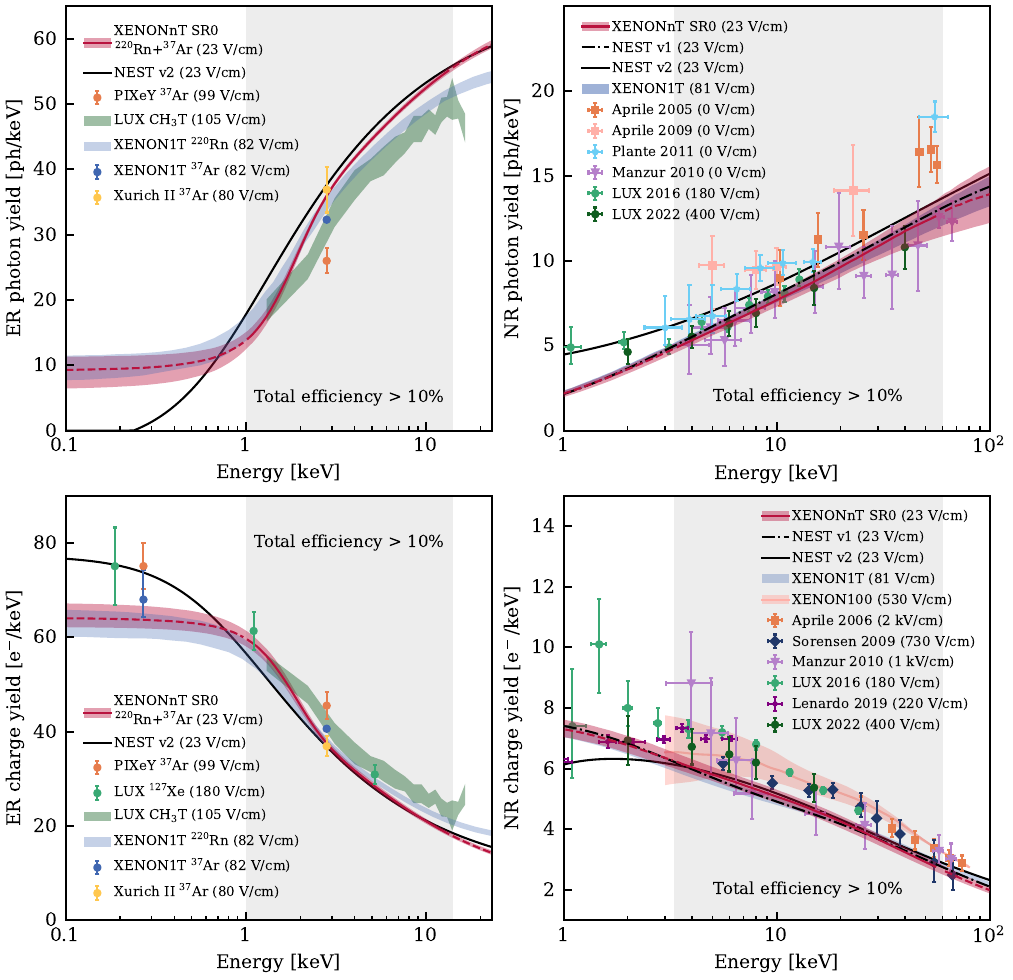}
    \caption{Photon (top) and charge (bottom) yields as functions of deposited energy for ER events (left) and NR events (right). Dark red lines and shaded bands show the XENONnT liquid xenon emission models with the best-fit parameters and their uncertainties, respectively. The fit results from XENON1T (blue shaded bands) \cite{xenon1t_2019_ap2,1terratum}, and the models from NEST v1 (black dash-dotted line) and NEST v2 (black solid line), both evaluated at 23\,V/cm, are shown as well. Several measurements from literature performed at various drift fields as labeled in the legends (ER yields from \cite{xenon1t_2019_ap2,  aprile2023low, Baudis:2021dsq, Boulton:2017hub, LUX:2017ojt, LUX:2015amk}, NR yields from Aprile 2005~\cite{Aprile:2005mt}, Aprile 2006~\cite{Aprile:2006kx}, Aprile 2009~\cite{Aprile:2008rc}, XENON100~\cite{XENON100:2013smi}, Plante 2011~\cite{Plante:2011hw}, Sorensen 2009~\cite{XENON10:2008wre}, Manzur 2010~\cite{Manzur:2009hp}, LUX 2016~\cite{LUX:2016ezw}, Lenardo 2019~\cite{Lenardo:2019vkn}, and LUX 2022~\cite{LUX:2022qxb}) are also shown. Note that for the ER yields, a considerable dependence on the electric drift field is expected, as visible in the literature data points. The gray shaded regions mark the energy ranges in which the total detection + selection efficiency in the WIMP analysis is above 10\,\%.}
    \label{fig:yields}
\end{figure*}

In the ER emission model, the parametrization is the same as in XENON1T \cite{xenon1t_2019_ap2} with very wide flat priors, referred to as free priors.
For the NR emission model in turn, the parameter priors are taken from ~\cite{Lenardo_2015}. 
A Gaussian-like distribution with different widths on either side is used as the prior of parameters with asymmetric uncertainties.
However, due to the low drift field of 23\,V/cm, where no literature data on NR yields exist, the validity of the field dependence in the model is unverified.
Therefore, the field dependence in the emission model from~\cite{Lenardo_2015} was modified.
Two scaling parameters were freed ($\alpha$ and $\gamma$ in \cref{eq:nr_ex_ion_ratio} and \cref{eq:nr_ti} in the appendix), and two field dependence parameters ($\zeta$ and $\delta$) were fixed to the reported best-fit values.
The widths of all other parameter priors were doubled in order to allow more freedom in the fit.
Because the dependence of NR signals on the drift field has been shown to be small (see also literature measurements shown in \cref{fig:yields}), we consider this a reasonable choice.

The fit results of all emission model parameters are summarized in \cref{tab:bbf_param_summary} in the appendix.
No significant tension between the posteriors and the priors of any parameter is found.
Suitable goodness-of-fit (GOF) tests are performed to assess whether the best-fit models adequately describe the ${}^{220}$Rn and AmBe calibration data.
Specifically, we employ binned Poisson likelihood $\chi^2$ tests in the $\mathrm{cS1}$--$\mathrm{cS2}$ space.
We adopt an equiprobable binning scheme using the \textit{GOFevaluation} package \cite{gofevaluation}, ensuring that the number of expected events under the best-fit model is the same in each bin, $\mu_{\mathrm{bin}}$.
In \cref{fig:ernrgofs} the results of the 2D $\mathrm{cS1}$--$\mathrm{cS2}$ Poisson likelihood $\chi^2$ test are shown for both the ER and the NR fit, overlaid with the ${}^{220}$Rn and the AmBe calibration data, respectively.
The resulting p-values of the tests indicate no significant discrepancy between the best-fit models and the calibration data.
The few outlier events below the NR band in the right panel are modeled by accounting for S2 charge losses near the TPC walls (see also \autoref{sec:appendix_detector_response_model} in the appendix).

\Cref{fig:yields} shows the photon and charge yields as a function of deposited energy for the ER and NR emission models using our best-fit parameters (dark red).
For comparison, NEST models calculated at a drift field of 23\,V/cm and several published measurements at different drift fields are shown.
For the NR yields, we also compare our results to the NEST v1 model at a drift field of 23\,V/cm (red dashed lines in \cref{fig:yields}) which uses the same parametrization as the model in this work (see \cref{sec:microphys}), showing very good agreement.

\begin{figure*}[htb]
    \centering
    \includegraphics[width=\textwidth]{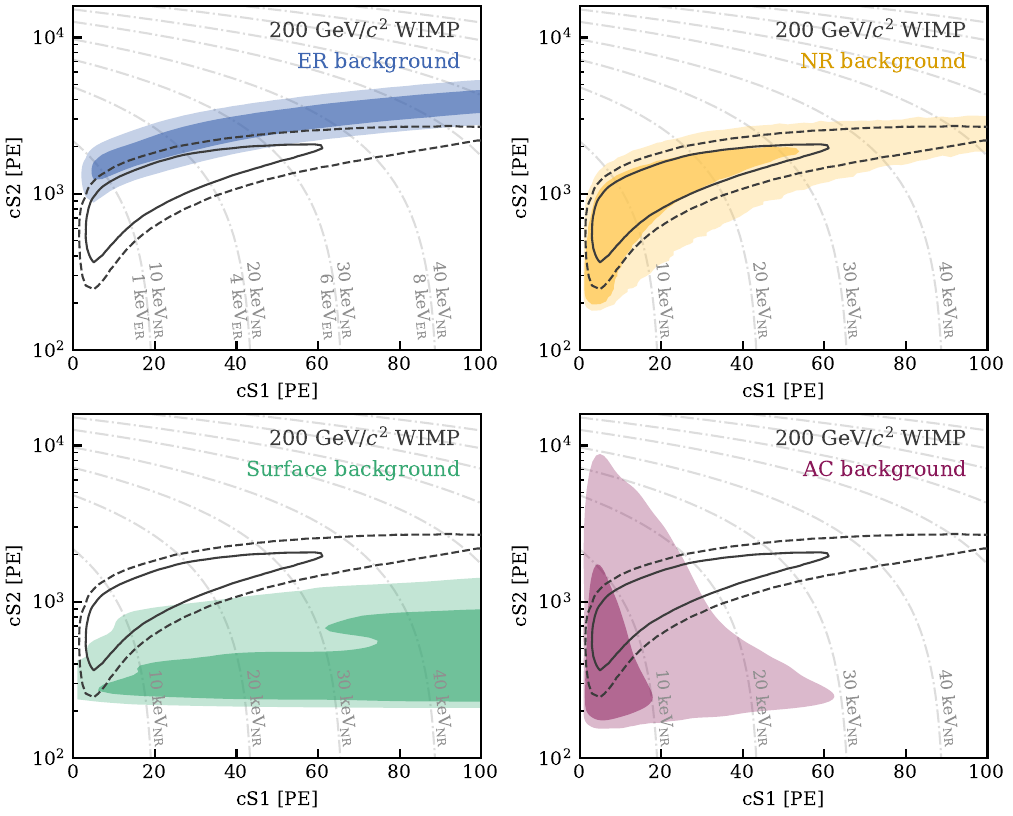}
    \caption{Distribution of each background component (colored) in $\mathrm{cS1}$--$\mathrm{cS2}$ space. The NR background includes both radiogenic neutron background and NR events by neutrinos. The probability density functions (PDFs) of backgrounds are shown as 1$\sigma$ (2$\sigma$) dark (light) regions, containing 68\% (95\%) expected events in the ROI. For reference, the contours are also shown for a spin-independent 200\,GeV$/c^2$ WIMP as a solid (dashed) dark gray line. The light gray dash-dotted lines are contours of constant NR-equivalent energy (and ER-equivalent energy in the top left panel) in units of $\mathrm{keV}$.}
    \label{fig:template_contours}
\end{figure*}

\section{Signal and background modeling}
\label{sec:signal_and_background}
The dominant sources of background in the WIMP search are ER events from intrinsic $\beta$-decays from materials and neutrinos (\cref{subsec:er_bkg}), NR events from radiogenic neutrons from detector materials (\cref{subsec:neutron_bkg}) and coherent neutrino scattering (\cref{subsec:cevns_bkg}), AC events (\cref{subsec:ac_bkg}), and surface background (\cref{subsec:surface_bkg}). Besides their total expected rates, the distribution of these background events in the analysis space cS1--cS2--\R is derived. In \cref{fig:template_contours}, the distributions of background models in the $\mathrm{cS1}$--$\mathrm{cS2}$ space are shown, compared to the signal model for elastic spin-independent scattering of a 200\,GeV$/c^2$ mass WIMP.

\subsection{WIMP Signal Model}
\label{subsec:wimp_sig}
\begin{figure}
    \centering
    \includegraphics[width=\columnwidth]{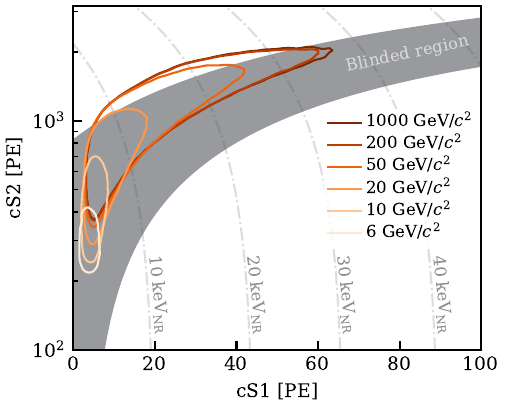}
    \caption{Expected distribution in cS1--cS2 space from spin-independent WIMP nuclear recoil for WIMP masses from 6\,GeV/$c^2$ to 1000\,GeV/$c^2$. The contours contain 68\% of expected events in the ROI. The blinded region, which is exclusively applied to the WIMP search dataset, is shown for reference as a gray band and dash-dotted lines are constant NR-equivalent energy contours.}
    \label{fig:wimp_template_fig}
\end{figure}

For non-relativistic, spin-independent elastic WIMP-nucleus coherent scattering, the event rate $R_{\mathrm{WIMP}}$ scales with the square of the atomic mass number $A^2$ of the target and can be written as \cite{LEWIN199687}
\begin{equation}
    \frac{\mathrm{d} R_{\mathrm{WIMP}}}{\mathrm{d} E} = \frac{\rho_\chi}{2m_\chi\mu_{\chi N}^2}\left\langle\frac{1}{v}\right\rangle\sigma A^2F^2(q),
\end{equation}
where $E$ is the recoil energy, $\rho_\chi$ is the local DM density of 0.3\,GeV/($c^2\,$cm$^3$), $m_\chi$ is the WIMP mass, $\mu_{\chi N}$ is the WIMP-nucleon reduced mass, $v$ is the WIMP velocity in the lab-frame, $\sigma$ is the WIMP-nucleon cross section, and $F(q)$ is the nuclear form factor as a function of the momentum transfer $q$ to the xenon nucleus \cite{helm}. The DM velocity distribution is averaged using the parameter values of the standard halo model with values from \cite{Lewin:1995rx,Smith:2006ym,McCabe:2013kea,Schoenrich:2009bx,Bland2016,gravity2021}, as recommended in \cite{Baxter_2021}.
\cref{fig:wimp_template_fig} shows the cS1--cS2 distribution of WIMP-nucleus scattering for different WIMP masses.
For increasing masses, the $68\%$ contours extend to higher cS1 and cS2 values, up to about 200\,GeV/$c^2$, where the shape does not change significantly anymore due to kinematic effects.
For this reason, confidence intervals in $\sigma$ (see \cref{sec:inference}) for $m_\chi\gtrsim200\,\mathrm{GeV}/c^2$ approximately scale proportional to the inverse of the assumed WIMP mass $m_\chi$.

For the spin-dependent WIMP-nucleus interaction, assuming natural abundances of xenon isotopes, only ${}^{129}$Xe and ${}^{131}$Xe can contribute since they are the only two stable isotopes of xenon with non-zero nuclear spin $J$. The cross section is usually written as
\begin{equation}
    \frac{\mathrm{d} \sigma_{\mathrm{SD}}}{\mathrm{d} q^2} = \frac{\sigma}{3\mu_{\chi N}^2v^2}\frac{\pi}{2J+1}S_N(q).
\end{equation}
The axial-vector structure factor of xenon $S_N$ is taken from \cite{klos2013large}. Note that the uncertainty on the structure factor dominates especially for the ``neutron-only'' case. In the XENONnT SR0 analysis, the mean of the structure factor is used.

Uncertainties from the posterior distribution of the NR model parameters and other efficiencies can be propagated to the NR rate uncertainty. For a 6\,GeV$/c^2$ (50\,GeV$/c^2$) WIMP, the relative rate uncertainty is $\sim$30\% (10\%). The NR model shape can also be affected by the posterior distribution. However, because of the low statistics of NR events in SR0 (for both background and signal expectation), the impact of the shape uncertainty of the NR model is negligible compared to the uncertainty of its absolute rate. Thus, the NR model shape uncertainty is not propagated to the final likelihood.

\subsection{Electronic Recoil Background Model}
\label{subsec:er_bkg}

\begin{figure}[htb]
    \centering
    \includegraphics[width=\columnwidth]{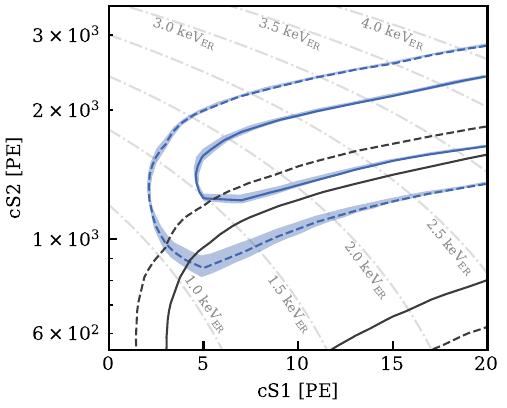}
    \caption{Shape uncertainty of the ER model constrained by the ${}^{220}$Rn calibration data. The blue (gray) solid and dashed lines represent the contours of ER background (200\,GeV$/c^2$ WIMP) containing 68\% and 95\% of expected events in the ROI. The transparent blue regions show the shape uncertainty by varying the two shape nuisance parameters by $\pm 1\sigma$. Dash-dotted lines are contours of constant ER-equivalent energy in units of $\mathrm{keV}_{\mathrm{ER}}$.}
    \label{fig:bbf_er_pca}
\end{figure}

ER events are one of the dominant background sources in the XENONnT SR0 WIMP search, primarily originating from $\beta$-decays of intrinsic radioactive contaminants such as $^{214}$Pb (a product of the $^{222}$Rn decay chain) and $^{85}$Kr. Contributions from ${}^{136}$Xe double-$\beta$ decays, solar neutrino interactions, and gamma events from the materials are also taken into account. In the ROI for the WIMP search, the total ER energy spectrum is approximately flat. The distribution of ER background events in $\mathrm{cS1}$--$\mathrm{cS2}$ space is generated from the ER model fitted to $^{220}$Rn and $^{37}$Ar calibration data, as discussed in \cref{sec:fit}. The total ER rate is fitted in the final WIMP likelihood, without applying an additional ancillary constraint in order to avoid potential systematic biases, which will be discussed in \cref{sec:inference}.

In contrast to the NR model, the uncertainty in the shape of the ER model can affect the sensitivity of the WIMP search. Ideally, the posterior distribution of all parameters should be propagated to the ER model as nuisance parameters in the WIMP search likelihood function. However, an excessive number of correlated nuisance parameters becomes computationally challenging. In XENONnT SR0, a principal component analysis (PCA)~\cite{pca} is used to remove correlation among parameters from the MCMC sampler. All ER parameters shown in \cref{tab:bbf_param_summary}, together with $W$, $g_1$, and $g_2$ (see details in the appendix), are included in the PCA. Different from the original PCA, the variance of 
\begin{equation}
\label{eq:pca_fom}
    t\equiv\sum_i \frac{s_i^2}{s_i+b_i}
\end{equation}
along each principal component is used to quantify its importance.
Here, $s_i$ and $b_i$ are the probability of a 50\,GeV$/c^2$ WIMP and ER background in the $i^{\text{th}}$ bin of cS1--cS2 space, respectively. A larger variance of $t$ means that the uncertainty of that component can be more impactful for the WIMP sensitivity. In the end, the two most important principal components are kept as nuisance parameters. These two shape parameters are constrained within the $^{220}$Rn ancillary term of the WIMP search likelihood function, as shown in \cref{fig:bbf_er_pca}. While adding an additional $^{37}$Ar ancillary term could further constrain the shape uncertainty at low energies, it was not included to maintain a conservative approach, as $^{220}$Rn already provides a sufficiently tight constraint on the shape uncertainty.

\subsection{Neutron Background Model}
\label{subsec:neutron_bkg}

\begin{figure*}[hbt]
\centering
\begin{minipage}[b]{.5\textwidth}
        \centering
        \includegraphics[width=\columnwidth]{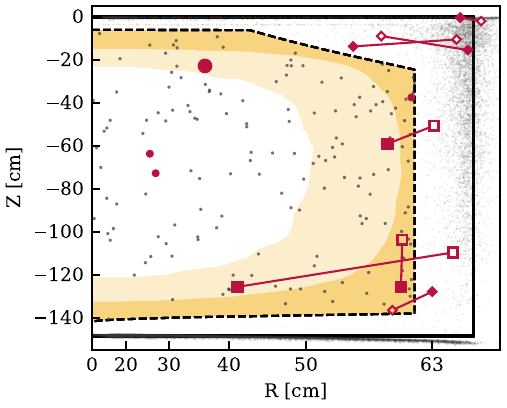}
\end{minipage}
\begin{minipage}[b]{.5\textwidth}
        \centering
        \includegraphics[width=\columnwidth]{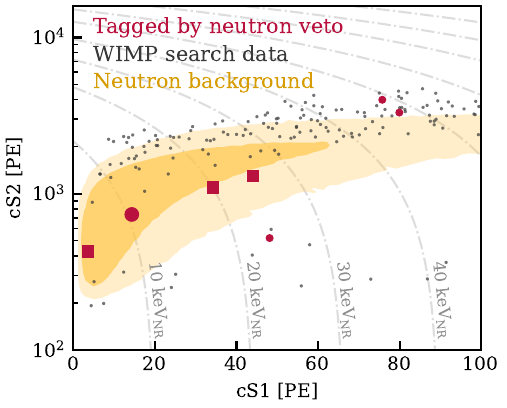}
\end{minipage}
\caption{Spatial (left) and cS1--cS2 (right) distribution of events tagged by the neutron veto (red markers). Black dashed and solid lines in the left plot indicate the FV and physical TPC boundary, respectively. The regions containing 68\% (95\%) of the total expected neutron background are indicated in dark (light) yellow. The neutron veto tagging yields four SS events (circles) and seven MS events, from which three are reconstructed with their largest S2 signals inside the FV (squares) and the other four outside (diamonds). Solid and hollow markers represent the reconstructed positions of the largest and the second-largest S2 signal, respectively. The right plot only shows events inside the FV; from these, the four events within the neutron background band (large markers; one SS and three MS) were used to estimate the neutron background rate. The other three SS events are in agreement with the prediction of falsely tagged ER and surface background events (c.f. \cref{fig:template_contours}). For reference, the events surviving all data selections, not tagged by the neutron veto, are displayed as gray dots. Also here, the right panel displays only those data points within the FV, which correspond to the unblinded WIMP search events.}
\label{fig:nr_background_rz_plot_with_ss}
\end{figure*}

Radiogenic neutrons are mainly produced through spontaneous fission and $(\alpha, n)$ reactions in the detector materials due to intrinsic traces of radioactive impurities. The cosmogenic neutron background is subdominant \cite{xenonnt_projection}, hence it is neglected in XENONnT SR0 analysis. The neutron yield and energy per isotope and material are calculated with the \textit{SOURCES-4A} software~\cite{osti_15215}, using the radioactive impurity levels of the relevant detector components obtained via the combined XENON1T and XENONnT radioassay campaigns~\cite{1t_screening_paper,aprile2022material}. A full-chain simulation pipeline~\cite{ramirez_thesis} is used to estimate the neutron background rate, the geometrical distribution, and the cS1--cS2 distributions for SR0.

The propagation of the neutrons in the XENONnT detector is simulated with the \textit{Geant4} toolkit~\cite{Agostinelli:2003geant4, ALLISON2016186}, where the recoil type and energy deposition per interaction in the target are recorded. We compute the number of photons and electrons for a given interaction via the custom-developed \textit{epix} package~\cite{epix}, which utilizes the energy-dependent LXe response derived from the SR0 calibrations, shown in \cref{fig:yields}. This package also handles the clustering of the individual energy-depositing steps at the LXe microphysics scale before the quanta generation. The photons and electrons produced by \textit{epix} are passed to the waveform simulator (\textit{WFSim})~\cite{wfsim}, which computes the S1 and S2 signals up to the waveform level, by means of a precise set of simulations- and data-driven corrections which characterize the XENONnT detector response. The event-by-event simulated waveforms share the same data structure as the science data after applying PMT and Data Acquisition (DAQ) effects, which allows us to process them with the same software used for the real data (\textit{straxen} \cite{joran_r_angevaare_2022_6854329}), as well as to apply the same data selections. The final SS neutron rate arises from weighting the rates obtained via the entire waveform simulation pipeline with the specific activities of the corresponding material and isotopic neutron yield.

When a neutron scatters in the $\sim$250\,kg of LXe between the cathode and the bottom PMT array, only the scintillation light for these events can be detected. The electrons, in turn, are lost due to an electric field pointing in the opposite direction to the active volume. Neutron events that consist of scatters above and below the cathode are referred to as neutron-X events, where ``X'' means additional S1 contribution from charge-insensitive scatters. They are modeled as a separate background since they have a larger cS1 to cS2 ratio than normal neutron scatters.

The event parameters having discrimination power
on MS interactions, such as S2 pulse shape and PMT hit patterns of S1, are matched and validated with calibrations, such that the relevant data quality selections can be applied to the simulation outputs. Notably, a validation of the multiple-to-single scatter ratio and total rate in the TPC of the AmBe calibration data was conducted, from where we obtained the systematic uncertainty associated with the accuracy of the full-chain simulations.

With this agreement between simulations and calibration data, we decided before unblinding the WIMP ROI to proceed with the sideband unblinding of the events tagged by the neutron veto. Initially, this confirmed the simulation-driven neutron background prediction.
However, a mistake in the definition of the neutron veto time window was found after the unblinding of the WIMP ROI. After fixing this issue, a mismatch was found, with the neutron background rate being larger than predicted in the ROI \cite{aprile_search_2023}. We therefore decided to constrain the neutron background rate in a purely data-driven way based on the aforementioned sideband unblinding with the correct neutron veto tagging window.
The results of the sideband unblinding are shown in \cref{fig:nr_background_rz_plot_with_ss}. Based on the multiple-to-single scatter ratio of 2.2, the $53\%$ neutron veto tagging efficiency, and the three observed MS events and one SS neutron event tagged by the neutron veto in the fiducial volume, a data-driven prediction of $1.1^{+0.6}_{-0.5}$  events was derived for the SR0 exposure. We used the simulation-driven ratio between normal neutron background and neutron+X events to estimate the neutron+X event rate shown in Tab.\,\ref{tab:likelihood_params}.

No further modification was propagated into the analysis after the data-simulations rate mismatch was identified: the $250\,\mathrm{\upmu s}$ time veto window between the TPC and the neutron veto, chosen due to the reduced background rate initially predicted, and the fiducial volume of the WIMP search remained as defined prior to unblinding. An underestimated contamination from some of the surrounding materials is considered as a possible cause of the discrepancy between the expected rate and the observed neutron background. Studies to constrain the material's radiopurity by means of the high-energy gamma rays for specific detector regions are ongoing.

\subsection{\cevns Background Model}
\label{subsec:cevns_bkg}
Neutrinos can also contribute to the NR background via coherent elastic neutrino-nucleus scattering (\cevns). Signals from solar, atmospheric, and diffuse supernova neutrinos (DSN) will be in the WIMP search ROI. Due to the weak interaction between neutrinos and nuclei, \cevns events are expected to be single-scatter and spatially uniform.

The recoil energy spectrum of solar \cevns is almost identical to that of a 6\,GeV$/c^2$ spin-independent WIMP \cite{xenon1t_cevns}, and the flux is $(5.25\pm0.20)\times10^{6}\,\mathrm{cm}^{-2}\mathrm{s}^{-1}$ \cite{SNO:2011hxd,BOREXINO:2018ohr}. After applying all selections and their corresponding efficiencies, the expected number of events is $(0.19\pm0.06)$ in SR0, also shown in \cref{tab:likelihood_params}. Atmospheric neutrinos and DSN mainly affect the search for heavier WIMPs. Their recoil energy spectra are taken from \cite{xenonnt_projection}, and the SR0 expectation value is $(0.05\pm0.02)$ events.
Due to the low cross section and the similarity to WIMP interactions, \cevns background will be the major limitation to the WIMP search sensitivity of the next generation of LXe experiments.

\subsection{Accidental Coincidence Background Model}
\label{subsec:ac_bkg}

AC events consist of incorrectly paired S1 and S2 signals.
These S1 and S2 signals can occur, for example, when either the S1 or the S2 signal of a physical event is not reconstructed due to detector effects, or when a single electron S2 signal is misclassified as an S1 signal.
Such signals are referred to as ``isolated'' S1 and S2 signals.
If an isolated S1 and an isolated S2 fall within the event-building time interval \cite{xenon1t_2019_ap1}, they form an AC event.

The AC background is modeled with a data-driven approach. Isolated S1 and S2 signals as well as their surrounding S1 and S2 signals that are close in time are sampled and paired into events. S1 signals $<$150\,PE are selected as isolated S1 signals, and the isolated S2 signals are taken from events whose S1 area is $<$150\,PE, together with all pulses in the event window. The AC event rate is computed as
\begin{equation}
    R_{\mathrm{AC}}=R_{\mathrm{isoS1}}\times R_{\mathrm{isoS2}}\times \Delta t,
\end{equation}
where $R_{\mathrm{isoS1}}$ and $R_{\mathrm{isoS2}}$ are the isolated S1 and S2 rates, and $\Delta t$ is the event-building time interval, which is defined according to the maximum drift time of 2.2\,ms in XENONnT.
Occasionally during SR0, some localized high rates of S2 signals appeared in the TPC.
Excluding these periods from the analysis, $R_{\mathrm{isoS1}}$ and $R_{\mathrm{isoS2}}$ remained stable throughout SR0 at an average rate of 1.5\,Hz and 80\,mHz, respectively.
In XENONnT, the isolated S2 rate is an order of magnitude higher compared to XENON1T~\cite{aprile_dark_2018}. This can be explained by the lower electron extraction efficiency which causes an increased rate of delayed electrons.
The isolated S1 and S2 signals are fed into the data processing pipeline \cite{jelle_aalbers_2022_6620274,joran_r_angevaare_2022_6854329} to reconstruct the events. This provides the background distribution of all relevant event properties, especially the distribution in cS1 and cS2, as shown in \cref{fig:template_contours} (lower right).

To suppress AC events, a selection criterion based on a gradient-boosted decision tree (GBDT) classifier utilizing the S2 pulse shape and the drift time was developed. Due to an insufficient model for the S2 pulse shape near the perpendicular wires, the GBDT classifier is only applied in the region at least 4.45\,cm far from the wires (far-wire region), while in the region near the wires (near-wire region), a data-driven S2 shape selection is applied instead. 
The resulting AC background rate per ton-year in the near-wire region is about 5.6 times higher compared to the rate in the far-wire region.
Consequently, the modeling of the TPC response is split into two parts.

\begin{figure}
    \centering
    \includegraphics[width=\columnwidth]{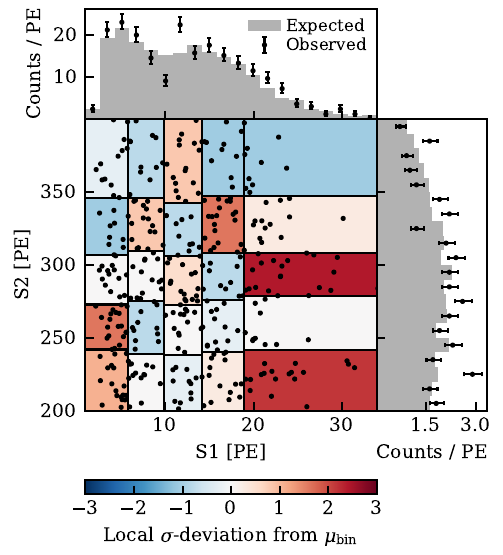}
    \caption{AC model validation with ${}^{37}$Ar calibration data in S2 versus S1 area. The AC model was cross-checked against this high-statistics dataset using a 2D GOF test evaluated in an equiprobable binning scheme. The color scale shows the deviation from the predicted number of events in each bin $\mu_\text{bin}=14.0$ in units of sigma. The projections to S1 and S2 are shown above and to the right of the 2D plot. No significant deviations were observed in the bulk of the distribution.}
    \label{fig:ac_37Ar}
\end{figure}

Validation performed on calibration data provides the rate and shape uncertainty of the AC model. For this, we used  dedicated AC-rich datasets, such as sidebands defined by inverting anti-AC selections. We compared the predicted and observed AC events by performing both 1D and 2D GOF tests with equiprobable bins in all relevant parameters.
The AC predictions are provided by the AC modeling method discussed above.
Because of the large statistics of ${}^{37}$Ar calibration data, it delivers the most stringent constraint on the AC model. ${}^{37}$Ar events with S2 areas smaller than 400\,PE are selected to test the AC model. The predicted number of AC events was 731.6, while 733 events were observed in the data. The statistics in the AC model is very large so the uncertainty of the predicted AC rate is neglected in the following GOF tests. The 1D GOF tests in S1, S2, \z, \R, and the 2D test in S1--S2 all yield p-values between 0.05 and 0.95. The result of the 2D test with a p-value of 0.61 is shown in \cref{fig:ac_37Ar}. The model was further validated with events removed by selections targeting the AC background inside the WIMP ROI.
Similar tests were also performed on ${}^{220}$Rn calibration data. All these tests show the model and data to be compatible.
Due to an over-fitting issue found in the GBDT training process, we conservatively estimated the rate uncertainty to be 30\% for both the near- and far-wire regions. The expectation value of AC events in the WIMP search dataset in the ROI is $(4.3\pm 0.9)$.

\subsection{Surface Background Model}
\label{subsec:surface_bkg}

\begin{figure}[htb]
    \centering
    \includegraphics[width=\columnwidth]{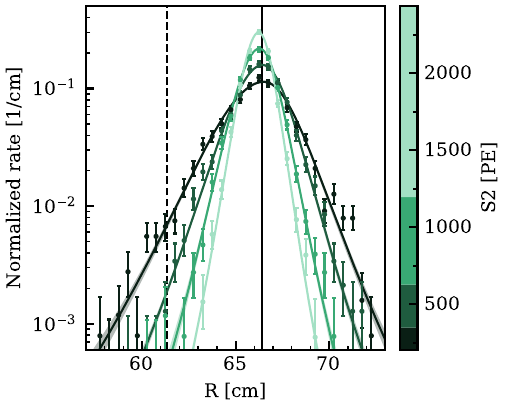}
    \caption{The radial surface background model is shown for four slices in S2 area. For each slice, a Student's generalized skew-t distribution is fitted to $^{210}$Po $\alpha$-events. The outer radius of the physical TPC as well as of the fiducial volume are shown as a solid and a dashed black line, respectively.}
    \label{fig:wall}
\end{figure}

In XENONnT, the TPC wall is made of Polytetrafluoroethylene (PTFE), which is known to accumulate isotopes from the decay chain of atmospheric $^{222}$Rn, which decays down to $^{210}$Pb, an isotope with a half-life of 22.2 years.
Radioactive decays on the wall surface can result in events with reduced S2 signals due to charge losses, which gives rise to a background that can leak into the WIMP signal region.
Three components contribute to the surface background in the WIMP ROI: the two $\beta$-decays of $^{210}$Pb and $^{210}$Bi, and the recoiling $^{206}$Pb following the $\alpha$-decay of $^{210}$Po.
Independent screening measurements indicate a $^{210}$Po $\alpha$-decay activity on the PTFE surface of $(20 \pm 3)\,\mathrm{mBq}\,\mathrm{m}^{-2}$~\cite{aprile2022material}.
Due to the complexity of electric field conditions near the surface and the loss of ionization electrons to the detector walls, we employ a data-driven approach to model this background component in the space of cS1--cS2--\radius--\z.

The surface background model in the space of \radius--\z was developed using $^{210}$Po $\alpha$-events. These events are mono-energetic (5.4 MeV) and thus easily identifiable through their characteristic S1 signal, yet they present a wide range of S2 sizes due to variable charge loss to the walls. The \radius and \z distributions of $^{210}$Po $\alpha$-events were seen to match those of the lower-energy $\beta$-events. The radial profiles were modeled using a Student's generalized skew-t distribution.
The radial distribution was fitted independently for different S2 sizes as shown in \cref{fig:wall} to account for the S2-size dependent position resolution \cite{nT_sr0_analysispaper_I}.
The \z profile was modeled using a linear function, to account for lower rates of surface events observed at the bottom of the TPC, likely due to the increased charge insensitivity near the walls at the bottom of the detector \cite{aprile_field_cage_2023}.

The cS1--cS2 distribution was modeled using a 2D Gaussian adaptive kernel density estimation (aKDE), built using events reconstructed outside the detector walls. The resulting model was then validated against the events reconstructed between the walls and the fiducial volume, in order to rule out any radial bias in the cS1--cS2 model. \Cref{fig:template_contours} (lower left) shows the projection of the four-dimensional model on cS1--cS2 in the ROI. The absolute rate of events in the blinded region was inferred from the radial distributions of adjacent, non-blinded regions in cS1--cS2.
 
Uncertainties in the model were obtained from the \radius and \z fit parameter uncertainties, as those are the parameters of interest in the development of the FV. Uncertainties on the overall measured surface event rate were also propagated. Uncertainties in the (cS1, cS2) aKDE were neglected, as toy-MC tests were performed to show that they had little impact on overall expectation.
For the FV, a maximum radius of \SI{61.35}{\cm} compared to the \SI{63}{\cm} used in \cite{aprile_low_er_search_2022} was chosen to reduce the number of surface events to $(14\pm3)$, which improves the robustness against mismodeling.

\section{Statistical Inference}
\label{sec:inference}
In this section, we describe the statistical methods used to derive the WIMP search results, which generally follow the recommendations formulated in \cite{Baxter_2021}.
We first describe the likelihood function used for the search in \cref{sec:likelihood} and its nuisance parameters in \cref{sec:nuisance_and_constraints}, followed by an illustration of the procedure for constructing confidence intervals and computing the discovery significance in \cref{sec:intervals}.
The GOF test to validate the best-fit models and the blinding procedure are described in \cref{sec:gof_blinding}.
We omit some details already presented in  \cite{xenon1t_2019_ap2}, as the methods here presented build upon that previous work.

\subsection{ Definition of the Likelihood Function}
\label{sec:likelihood}
The fundamental ingredient for the statistical analysis of our WIMP search data is the likelihood function $\mathcal{L}(\sigma, \boldsymbol{\theta})$.
It depends on the WIMP-nucleon cross section $\sigma\geq 0$, which is our parameter of interest, and a set of nuisance parameters $\boldsymbol{\theta}$.
The likelihood function is defined as a product of four terms: two for the WIMP search dataset ($\mathcal{L}_\text{near-wire}$ and $\mathcal{L}_\text{far-wire}$), one for the $^{220}$Rn calibration dataset ($\mathcal{L}_\text{cal}$), and one for ancillary measurements of nuisance parameters ($\mathcal{L}_\mathrm{anc}$):

\begin{equation}
\begin{aligned}
    \mathcal{L}(\sigma, \boldsymbol{\theta}) = {} & \mathcal{L}_\text{near-wire}(\sigma, \boldsymbol{\theta}) \\
    & \times \mathcal{L}_\text{far-wire}(\sigma, \boldsymbol{\theta}) \\
    & \times \mathcal{L}_\mathrm{cal}(\boldsymbol{\theta}) \\
    & \times \mathcal{L}_\mathrm{anc}(\boldsymbol{\theta}).
    \label{eq:total_likelihood}
\end{aligned}
\end{equation}

In the WIMP search data, we categorize events based on their proximity to the perpendicular wires of the gate and anode electrodes, distinguishing between those that are near ($\leq$\SI{4.45}{cm}, corresponding to about $17\%$ of the total FV) and far from the wires in the \x--\y plane. This approach allows us to account for the higher AC rate observed near the wires as discussed in \cref{subsec:ac_bkg}, without introducing a full position-dependence in the likelihood.
Other backgrounds, in particular radiogenic neutron and surface background, exhibit a substantial radial dependence. Thus, the likelihood is modeled and evaluated in \radius, in addition to cS1 and cS2. For the near-wire region, the radial component is not included in the modeling.
Each WIMP search term is an extended unbinned likelihood function of the form

\begin{equation}
\begin{aligned}
    \mathcal{L}_\text{region}(\sigma, \boldsymbol{\theta}) = {} &\mathrm{Pois}(N|\mu_\mathrm{tot}(\sigma, \boldsymbol{\theta})) \\
    &  \times\prod_{i=1}^N\left[\sum_c\frac{\mu_c(\sigma, \boldsymbol{\theta})}{\mu_\mathrm{tot}(\sigma, \boldsymbol{\theta})} \times f_c(\vec{x}_i|\boldsymbol{\theta})\right],
\end{aligned}
\label{eq:science_likelihood}
\end{equation}
where the index ``region'' runs over ``near-wire'' and ``far-wire''.
The index $i$ runs over all $N$ observed events $\vec{x}_i$ in the WIMP search far-wire (near-wire) dataset, where $\vec{x}_i$ is a vector with entries cS1, cS2, and \radius (cS1 and cS2).
The PDFs $f_c$ of each signal and background component $c$ with expectation values $\mu_c$ are evaluated for each event.
The total expectation value is given by $\mu_\mathrm{tot}(\sigma, \boldsymbol{\theta}) = \sum_c \mu_c(\sigma, \boldsymbol{\theta})$.

The term $\mathcal{L}_\mathrm{cal}(\boldsymbol{\theta})$ is the extended unbinned likelihood function of the $^{220}$Rn calibration dataset.
It depends on the two shape parameters introduced in \cref{subsec:er_bkg} that parameterize the range of models consistent with the ER calibration data, selected using the PCA method.
By incorporating this likelihood term, we simultaneously fit the shape parameters to the $^{220}$Rn calibration dataset and the ER background model in the WIMP search dataset.
Through this procedure, the constraint from the calibration data on the shape of the ER model is propagated to the final inference.

Finally, the ancillary likelihood function $\mathcal{L}_\mathrm{anc}(\boldsymbol{\theta})$ is a product of constraint terms for some of the nuisance parameters $\boldsymbol{\theta}$, which are detailed in the following section.

\subsection{Nuisance Parameters and Constraints}
\label{sec:nuisance_and_constraints}

\begin{table*}[ht]
  \centering
  \caption{Parameters of the XENONnT SR0 WIMP search likelihood function and their constraints. The right column shows best-fit values with an unconstrained \SI[per-mode=symbol]{200}{\giga\eVperc\squared} WIMP signal component (spin-independent coupling). For the rate parameters, all values are given in units of ``events in the SR0 exposure''. Near and far-wire refers to the region in the \x--\y plane near and far from the perpendicular wires. In the near-wire region, which constitutes approximately $17\%$ of the total FV, the AC expectation value is comparable to the one in the far-wire region due to the higher AC rate near the wires.}
  \begin{tabular}{llrr}\toprule
  Rate Parameter    & Constraint  & Nominal & Best Fit\\\midrule
  ER    & WIMP search data & \expectationernominalbothall & \expectationerbestbothall\\
  Neutron & Ancillary measurement & \expectationradiogenicnominalbothall & \expectationradiogenicbestbothall\\
  Neutron-X & Ancillary measurement & \expectationradiogenicXnominalbothall & \expectationradiogenicXbestbothall\\
  CE$\nu$NS (Solar $\nu$)& Ancillary measurement & \expectationcevnsnominalbothall & \expectationcevnsbestbothall\\
  CE$\nu$NS (Atm + DSN) & Ancillary measurement & \expectationatnunominalbothall & \expectationatnubestbothall\\
  AC (near-wire)   & Ancillary measurement & \expectationACwirenominalonall & \expectationACwirebestonall\\
  AC (far-wire)   & Ancillary measurement & \expectationACnonwirenominaloffall & \expectationACnonwirebestoffall\\
  Surface    & Ancillary measurement & \expectationwallnominalbothall & \expectationwallbestbothall\\
  $^{220}$Rn calibration & $^{220}$Rn dataset & \expectationercalibrationnominalboth & \expectationercalibrationbestboth
    \\\\
  \toprule
  Shape Parameter & Constraint  & Nominal & Best Fit\\\midrule
  ER shape parameter 1 & $^{220}$Rn dataset & \expectationparonenominal & \expectationparonebest\\
  ER shape parameter 2 & $^{220}$Rn dataset & \expectationpartwonominal & \expectationpartwobest\\

  \\\\
  \toprule
  Signal Parameter     & Constraint  & Nominal & Best Fit\\\midrule
  Relative signal efficiency  & NR model uncertainty & $1.00\pm0.09$ & $1.0$\\
  WIMP cross section [$10^{-47}\mathrm{cm^{-2}}$]&  & Parameter of interest & 3.22\\
  \bottomrule
  \end{tabular}
  \label{tab:likelihood_params}
\end{table*}

In addition to our parameter of interest $\sigma$, we parameterize the space of background hypotheses using nuisance parameters $\boldsymbol{\theta}$ in the likelihood function of \cref{eq:total_likelihood}.
These parameters control both background rates and shape.
The relative signal efficiency is  a special parameter that quantifies the rate uncertainty of the NR emission model, as described in \cref{subsec:wimp_sig}.
By construction, it takes the nominal value of 1.0 for an unconstrained fit, but it can depart from this value for fits with constrained cross sections, leading to a broadening of the confidence interval.
In total, twelve nuisance parameters are defined, which are listed in \cref{tab:likelihood_params}.
The uncertainties of the nominal expectation values correspond to the width of a Gaussian constraint term.
The ancillary likelihood function $\mathcal{L}_\mathrm{anc}(\boldsymbol{\theta})$ is the product of all constraint terms. 
Most parameters are constrained via ancillary measurements, which are obtained from sidebands (e.g. data outside the ROI) or external measurements in combination with simulations.
More details on the constraints on neutron, \cevns, AC, and surface background rates were discussed in \cref{sec:signal_and_background}.
The ER rate is a free parameter in the fit and is fully constrained by the WIMP search data.
The ER shape parameters obtained with PCA are constrained via the simultaneous fit of the \rntwotwozero calibration dataset via the term $\mathcal{L}_\mathrm{cal}(\boldsymbol{\theta})$ in \cref{eq:total_likelihood}.

The best-fit values from the XENONnT SR0 WIMP search data \cite{aprile_search_2023} for a fit including an unconstrained \SI[per-mode=symbol]{200}{\giga\eVperc\squared} signal component are given in the last column of \cref{tab:likelihood_params}.
The corresponding uncertainties represent the two-sided one-sigma confidence intervals derived from a profile likelihood scan in the respective parameter.
More details on the construction of confidence intervals are discussed in the following.

\subsection{Confidence Intervals and Discovery Significance}
\label{sec:intervals}
The construction of confidence intervals is based on the profile likelihood test statistic
\begin{equation}
    q(\sigma)\equiv -2 \ln\frac{\mathcal{L}(\sigma, \boldsymbol{\hat{\hat{\theta}}})}{\mathcal{L}(\hat{\sigma}, {\boldsymbol{\hat{\theta}}})}.
    \label{eq:plr}
\end{equation}
Quantities with a single hat denote the global maximum likelihood estimator of a parameter, while $\boldsymbol{\hat{\hat{\theta}}}$ denotes the set of nuisance parameters that maximize the likelihood if the cross section is fixed to a given $\sigma$. 
The likelihood is defined for a specific signal model, for example a \wimp with a certain mass and interaction type (e.g. spin-independent or spin-dependent coupling).
In our statistical inference, we consider signal models across a range of WIMP masses, from \SI[per-mode=symbol]{6}{\giga\eVperc\squared} to \SI[per-mode=symbol]{500}{\giga\eVperc\squared}.
For higher WIMP masses, the PDF remains constant and the flux for a given cross section decreases approximately linearly with mass, as discussed in \cref{subsec:wimp_sig}.
According to this, limits can be extrapolated even beyond \SI[per-mode=symbol]{500}{\giga\eVperc\squared}.

Knowing the distribution of $q(\sigma)$ under different hypotheses is essential for calculating discovery significances and confidence intervals.
Due to the low rate of background resembling the signal, asymptotic formulae as discussed in \cite{Cowan2011}  are not necessarily a good approximation.
Therefore, we primarily use \toymcs to estimate the distribution $g(q(\sigma) | \sigma)$ of the test statistic given that $\sigma$ is the true cross section.
We use the custom-developed \textit{blueice} \cite{blueice} framework to define the likelihood function. This Python package provides an efficient interpolation (``template morphing'') between PDFs evaluated for different discrete values of shape nuisance parameters.
This allows shape parameters to be considered for which we have no analytical description.

\begin{figure}
    \centering
    \includegraphics[width=\columnwidth]{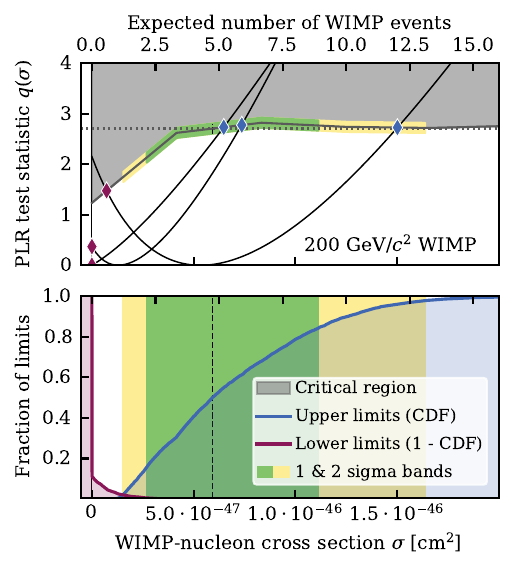}
    \caption{Illustration of the confidence interval construction and distribution of limits on the WIMP-nucleon cross section $\sigma$ of a \SI[per-mode=symbol]{200}{\giga\eVperc\squared} WIMP for background-only \toymcs. Top: The test statistic $q(\sigma)$ as a function of the cross section $\sigma$ is shown for three \toymcs. The intersections with the threshold of the critical region (gray line) yield the 90 $\%$ CL upper (blue diamonds) and lower limits (purple diamonds). Bottom: (Inverse) cumulative distribution function (CDF) of upper (lower) limits for background-only \toymcs. The bands containing 68\,$\%$ (green) and 95\,$\%$ (yellow) of ULs, as well as the median UL (dashed line), are indicated. Note the shared abscissa with the cross section (bottom) and the corresponding expected number of signal events (top).}
    \label{fig:profile_construction_limits}
\end{figure}

For each signal model considered, we compute the discovery significance and the confidence interval by comparing the measured test statistic $q(\sigma)$ with the distribution under each hypothesis $g(q(\sigma)|\sigma)$ \cite{feldman_cousins_1998,PDG}.
Testing $\sigma=0$ yields the (local) discovery significance, while confidence intervals are computed by finding the region where $\sigma$ is rejected at a $90\%$ confidence level given the data, which is illustrated for three background-only \toymcs (three black parabola-like lines) in \cref{fig:profile_construction_limits} (top).
The minimum of each curve corresponds to the respective best-fit value $\hat{\sigma}$.
Upper limits (ULs) and lower limits are obtained by finding the cross sections at which $q(\sigma)$ reaches the critical region.
The threshold of the 90 $\%$ CL critical region is precomputed as the 90th percentile of the test statistic distribution from \toymcs \textit{with} injected signal corresponding to the respective cross section.
For low cross sections, it deviates from the asymptotic two-sided threshold indicated as a horizontal gray dotted line.

Repeating this procedure $\mathcal{O}(1000)$ times yields the expected distribution of ULs that can be obtained in the absence of any signal, shown in \cref{fig:profile_construction_limits} (bottom).
The experiment's sensitivity is given by the median of these ULs and the sensitivity band (``Brazil band'') marked by the $\pm$2-sigma and $\pm$1-sigma quantiles indicates the spread.
The distribution of lower limits for background-only \toymcs is also shown in \cref{fig:profile_construction_limits} (bottom).
As recommended in \cite{Baxter_2021}, we decide on a discovery significance threshold for reporting two-sided confidence intervals of $3$ sigma, following~\cite{xenon1t_2019_ap2}.

\begin{figure}
    \centering
    \includegraphics[width=\columnwidth]{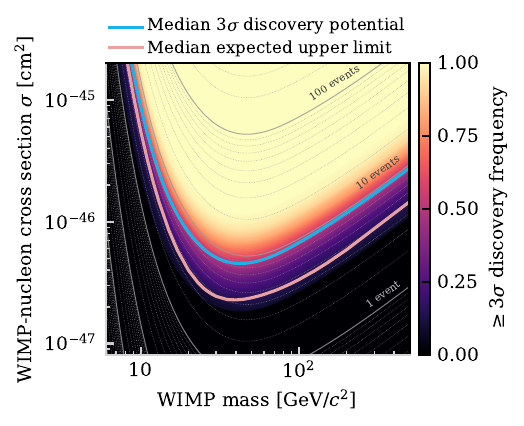}
    \caption{Three-sigma discovery potential as a function of WIMP mass and cross section. The blue line marks the 50\,\% discovery frequency, while the pink line represents the median expected upper limit from \cite{aprile_search_2023}. Thin contour lines indicate regions with an equal number of signal events.}
    \label{fig:discovery_potential}
\end{figure}

\Cref{fig:discovery_potential} shows the three-sigma discovery potential -- the expected frequency of achieving a discovery with at least three-sigma significance -- as a function of WIMP mass and cross section. 
The asymptotic threshold for a three-sigma discovery was applied.
Contours indicating equal signal event counts in the nominal model reveal that the median three-sigma discovery potential corresponds to approximately 10 events at high WIMP masses, decreasing to about 6 events at lower masses.

Statistical fluctuations as well as mismodeling, such as overestimated background rates, can yield arbitrarily low ULs, which may result in the spurious exclusion of models beyond the experiment's sensitivity.
To mitigate this issue, various methods have been proposed \cite{Read_2002,Kashyap_2010,cowan2011powerconstrained}.
The XENONnT SR0 WIMP search follows the recommendation of \cite{Baxter_2021} to use power-constrained limits (PCLs) \cite{cowan2011powerconstrained}.
In this method, a signal size threshold is selected, at which the experiment has a ``rejection power'' $M_\mathrm{min}$ -- the probability of excluding a given signal under the background-only hypothesis.
If an UL falls below this threshold, the threshold value is reported instead.
These thresholds correspond to the quantiles of the UL distribution used to compute the sensitivity band illustrated in \cref{fig:profile_construction_limits}.
For instance, choosing $M_\mathrm{min}=0.16$ sets the threshold to the -1 sigma quantile of the band, while $M_\mathrm{min}=0.5$ truncates ULs at the band's median.
In \cite{Baxter_2021}, the fiducial choice of $M_\mathrm{min}=0.16$ was suggested.
However, the choice was erroneously based on the \textit{discovery} power, which corresponds to the probability of rejecting the background-only hypothesis given an alternative hypothesis with a specific signal size.
In the absence of an updated recommendation, we chose the very conservative choice of $M_\mathrm{min}=0.5$ for the first WIMP search of XENONnT.

\subsection{Goodness of Fit and Blinding Procedure}
\label{sec:gof_blinding}
To verify that the best-fit models describe our WIMP search data well, we defined a GOF test before unblinding, and after studies to confirm the power to reject various forms of mismodeling.
The test uses a binned Poisson likelihood $\chi^2$ with 15 equiprobable bins in the cS1--cS2 space, defined from the model being tested.
To compute a p-value, the distribution of the $\chi^2$ test statistic under the best-fit hypothesis is derived through \toymcs.
Specifying the test and acceptance threshold (90\% CL) before unblinding ensured that the results were not influenced by statistical fluctuations expected from the low-statistic WIMP search dataset.
For the background-only fit we found a p-value of $0.67$ and for a fit with an additional \SI[per-mode=symbol]{200}{\giga\eVperc\squared} WIMP signal the p-value is $0.63$.
Both indicate no strong mismodeling in the predefined parameter space.
Using GOF tests with well-studied power to discover signal-like mismodeling was also designed to replace the ``safeguard'' ER shape parameter defined in  \cite{Priel_2017}, due to its computational cost and susceptibility to some kinds of mismodeling.

When searching for new signals in data, it is crucial to avoid the experimenter's bias on the result \cite{Klein2005_blind}.
In our WIMP search, we adopt a common strategy for bias mitigation: blinding the signal region of the WIMP search dataset until all steps of the data analysis are finalized.
Initially, events in the ER and NR bands of the WIMP search dataset with a reconstructed ER energy below $20\,\mathrm{keV_{ER}}$ were blinded.
This involved defining the ER and NR bands in cS1 and cS2 based on quantiles in cS2 for slices in cS1 of calibration data.
In the first step, all events above $10\,\mathrm{keV_{ER}}$ and those above the $-2$ sigma quantile of the ER band were unblinded for the analysis presented in \cite{aprile_low_er_search_2022}.
The signal region for the WIMP search (indicated in \cref{fig:wimp_template_fig}) remained blinded until the analysis procedure was finalized.
The unblinded SR0 WIMP search data showed no significant excess for any of the tested WIMP masses with local discovery p-values $\geq 0.2$.
For this reason, we reported new ULs on the spin-independent WIMP-nucleon cross section across WIMP masses ranging from \SI[per-mode=symbol]{6}{\giga\eVperc\squared} to \SI[per-mode=symbol]{500}{\giga\eVperc\squared}, with a minimum of $2.58 \times 10^{-47}\;\mathrm{cm}^2$ at \SI[per-mode=symbol]{28}{\giga\eVperc\squared}.

\section{Summary}
\label{summary}

The XENONnT WIMP dark matter search relies on a detector response model as well as simulation- and data-driven background models.
These were combined to construct a statistical model in cS1, cS2, and $\mathrm{R}$, which was used to infer limits on WIMP-nucleus scattering cross sections.

The full detector response model for electronic recoil (ER) and nuclear recoil (NR) interactions, including both the xenon emission and detector reconstruction model, was successfully fitted to calibration data.
Accurate simulations of particle interactions up to the data acquisition waveform level made this possible, in particular, to correctly model the S2 multiplicity of events with several, potentially unresolved energy deposits.
Using these models, we derived the distributions for ER and NR backgrounds, as well as the signals, in our analysis space.
Except for ER, the background rates were constrained with ancillary measurements.
The radiogenic neutron background rate was constrained by first matching the simulated ratio of multiple-to-single scatter interactions and the neutron veto tagging efficiency with NR calibration data. After unblinding the neutron veto-tagged events, these three inputs were combined to derive a prediction for the remaining non-vetoed single-scatter neutron background.
Two shape parameters were propagated to the final inference to account for the uncertainty of the ER model in cS1--cS2 space.
Accidental coincidence and surface background were modeled with data-driven approaches.
The validity of the models was confirmed in calibration data as well as science data outside the region of interest of the WIMP search.

A blind analysis was performed for the first science data from XENONnT.
The statistical methods largely follow the previous XENON1T approach and community recommendations, by using a \toymc-calibrated profile log-likelihood ratio test statistic.
One departure from these recommendations was raising the minimal required power of the power-constrained limit (PCL) threshold from 0.15 to 0.5, corresponding to placing upper limits only at or above the median unconstrained upper limit.
Both the calibration fits and the final best-fit model were assessed and found acceptable with goodness-of-fit tests that were chosen based on their mismodeling rejection power, and defined prior to unblinding the data. 
Analysis of the data with an exposure of 1.1 tonne-years revealed no signal excess over backgrounds.
Therefore, new upper limits on the spin-independent WIMP-nucleon cross section were derived.

\begin{acknowledgments}
We gratefully acknowledge support from the National Science Foundation, Swiss National Science Foundation, German Ministry for Education and Research, Max Planck Gesellschaft, Deutsche Forschungsgemeinschaft, Helmholtz Association, Dutch Research Council (NWO), Fundacao para a Ciencia e Tecnologia, Weizmann Institute of Science, Binational Science Foundation, Région des Pays de la Loire, Knut and Alice Wallenberg Foundation, Kavli Foundation, JSPS Kakenhi, JST FOREST Program, and ERAN in Japan, Tsinghua University Initiative Scientific Research Program, Domaine d’intérêt majeur en astrophysique et conditions d’apparition de la vie+ (DIM-ACAV+) Région Ile-de-France, and Istituto Nazionale di Fisica Nucleare.
This project has received funding/support from the European Union's Horizon 2020 research and innovation program under the Marie Skłodowska-Curie grant agreement No 860881-HIDDeN.

We gratefully acknowledge support for providing computing and data-processing resources of the Open Science Pool and the European Grid Initiative, in the following computing centers: the CNRS/IN2P3 (Lyon - France), the Dutch national e-infrastructure with the support of SURF Cooperative, the Nikhef Data-Processing Facility (Amsterdam - Netherlands), the INFN Centro Nazionale Analisi Fotogrammi (CNAF) (Bologna - Italy), the San Diego Supercomputer Center (San Diego - USA) and the Enrico Fermi Institute (Chicago - USA). We acknowledge the support of the Research Computing Center (RCC) at The University of Chicago for providing computing resources for data analysis.

We are grateful to Laboratori Nazionali del Gran Sasso for hosting and supporting the XENON project.
\end{acknowledgments}

\bibliography{references}

\appendix

\section{Details on the ER Emission Model}
\label{sec:appendix_er_emission_model}
Let $E$ be the recoil energy of an ER event.
The total number of detectable quanta $N_{\mathrm{q}}$ is sampled from a normal distribution
\begin{equation}
    N_{\mathrm{q}}\sim\mathrm{Norm}(E/W, \sqrt{fE/W}).
\end{equation}
$W$ is the mean energy needed to generate a quantum, and $f$ is the Fano factor of $~0.059$ \cite{Aprile:2009dv,DOKE1976353}.
The ER energy deposit will produce both excitons and electron-ion pairs.
Using the mean ratio between number of excitons and ions $\left\langle N_{\mathrm{ex}}/N_{\mathrm{i}}\right\rangle$, we simulate their numbers from a binomial distribution:
\begin{equation}
    N_{\mathrm{i}}\sim\mathrm{Binom}\left(N_{\mathrm{q}}, \frac{1}{1+\langle N_{\mathrm{ex}}/N_{\mathrm{i}}\rangle}\right),
\end{equation}
\begin{equation}
    N_{\mathrm{ex}}=N_{\mathrm{q}}-N_{\mathrm{i}}.
\end{equation}
A fraction $r$ of ions recombine with electrons, depending on the electric drift field $F$.
We parametrize the mean value of $r$ in the same way as in XENON1T \cite{xenon1t_2019_ap2}:
\begin{equation}
    \langle r\rangle=\frac{1}{\mathrm{e}^{-(E-q_0)/q_1}+1}\left(1-\frac{\log(1+\langle N_{\mathrm{i}}\rangle\varsigma)}{\langle N_{\mathrm{i}}\rangle\varsigma}\right),
\end{equation}
where
\begin{equation}
    \varsigma = \frac{1}{4}\gamma\mathrm{e}^{-E/\omega}F^{-\delta}.
\end{equation}
The fluctuation of $r$ is parametrized via
\begin{equation}
    \Delta r = q_2(1-\mathrm{e}^{-E/q_3}),
\end{equation}
and the true recombination fraction $r$ is then sampled from
\begin{equation}
    r \sim \mathrm{Norm}(\langle r\rangle, \Delta r).
\end{equation}
All fitted ER parameters are listed in \cref{tab:bbf_param_summary}.
Finally, the numbers of produced electrons and photons are given by
\begin{equation}
    N_{\mathrm{e}} \sim \mathrm{Binom}(N_{\mathrm{i}}, 1-r),
\end{equation}
\begin{equation}
    N_\gamma = N_{\mathrm{q}} - N_{\mathrm{e}}.
\end{equation}

\section{Details on the NR Emission Model}
\label{sec:appendix_nr_emission_model}
Let $E$ be the recoil energy of an NR event.
The total number of produced quanta $N$ is
\begin{equation}
    N\sim\mathrm{Norm}(E/W, \sqrt{fE/W}).
\end{equation}
In contrast to electron recoils, recoiling xenon nuclei lose a significant amount of their recoil energy via atomic motion and collisions with other xenon atoms, which are processes that are undetectable in a LXe TPC.
This quanta loss is modeled following the Lindhard theory~\cite{osti_4153115}, with the so-called Lindhard quenching factor $L$, such that the number of detectable quanta becomes
\begin{equation}
    N_{\mathrm{q}}\sim\mathrm{Binom}(N, L).
\end{equation}
Following the parametrization in~\cite{Lenardo_2015}, the Lindhard factor is given by
\begin{equation}
    L = \frac{\kappa\,g(\epsilon)}{1 + \kappa\,g(\epsilon)},
\end{equation}
where $g(\epsilon)$ is a function of deposited energy via
\begin{equation}
    g(\epsilon) = 3 \epsilon^{0.15} + 0.7 \epsilon^{0.6} + \epsilon,
\end{equation}
\begin{equation}
    \epsilon = 11.5\,Z^{-7/3} (E/\mathrm{keV}),
\end{equation}
with the nuclear charge number $Z=54$ for xenon.
The numbers of produced ions and excitons are then simulated by
\begin{equation}
    N_{\mathrm{i}}\sim\mathrm{Binom}\left(N_{\mathrm{q}}, \frac{1}{1+\langle N_{\mathrm{ex}}/N_{\mathrm{i}}\rangle}\right),
\end{equation}
\begin{equation}
    N_{\mathrm{ex}}=N_{\mathrm{q}}-N_{\mathrm{i}},
\end{equation}
with the exciton-to-ion ratio parametrized as 
\begin{equation}\label{eq:nr_ex_ion_ratio}
    \langle N_{\mathrm{ex}}/N_{\mathrm{i}}\rangle = \alpha F^{-\zeta} (1 - \mathrm{e}^{-\beta \epsilon}).
\end{equation}
Unlike ERs, the recombination fluctuation in NRs is usually small, thus the number of photons produced from the recombination of electron-ion pairs is
\begin{equation}
    N_\gamma^{\mathrm{re}} \sim \mathrm{Binom}(N_\mathrm{i}, r),
\end{equation}
where $r$ follows the Thomas-Imel box model \cite{PhysRevA.36.614}
\begin{equation}
    r = 1 - \frac{\log(1 + N_i \varsigma)}{N_i \varsigma},
\end{equation}
\begin{equation}\label{eq:nr_ti}
    \varsigma = \gamma F^{-\delta}.
\end{equation}
The number of excitons is further reduced by bi-excitonic and Penning quenching effects \cite{MEI200812}, such that the number of photons produced from de-excitation becomes
\begin{equation}
    N_\gamma^{\mathrm{de}} \sim \mathrm{Binom}(N_{\mathrm{ex}}, f_l),
\end{equation}
with the scintillation quenching factor
\begin{equation}
    f_l = \frac{1}{1 + \eta \epsilon^\lambda}.
\end{equation}
Then the final numbers of photons and electrons are
\begin{equation}
    N_\gamma = N_\gamma^{\mathrm{de}} + N_\gamma^{\mathrm{re}},
\end{equation}
\begin{equation}
    N_{\mathrm{e}} = N_{\mathrm{i}} - N_\gamma^{\mathrm{re}}.
\end{equation}

\section{Details on the Detector Response Model}
\label{sec:appendix_detector_response_model}

The simulations of $\mathrm{S1}$ and $\mathrm{S2}$ signals from $N_{\gamma}$ and $N_{\mathrm{e}}$ are almost independent of each other.

To simulate $\mathrm{S1}$ signals, we first introduce the spatially dependent scintillation gain $\Tilde{g_1}$ with DPE excluded:
\begin{equation}
    \Tilde{g_1}(\x,\y,\z) = \frac{g_1}{1+p_{\mathrm{DPE}}} \cdot \mathrm{S1corr}^{-1}(\x,\y,\z),
\end{equation}
where $g_1$ is the averaged scintillation gain, $\mathrm{S1corr}$ is the relative spatial correction factor, and $p_{\mathrm{DPE}}$ is the probability of the double photoelectron emission effect \cite{Faham_2015}.
Then, for a given number of photons $N_\gamma$ and the position of the recoil $\x$, $\y$, $\z$, the number of photons detected by the PMTs is
\begin{equation}
    N_{\mathrm{PhD}} \sim \mathrm{Binom}(N_\gamma, \Tilde{g_1}(\x,\y,\z)).
\end{equation}
Now accounting for the DPE effect, the expected number of photoelectrons (PE) in the S1 signal is simulated via
\begin{equation}
    N_{\mathrm{DPE}} \sim \mathrm{Binom}(N_{\mathrm{PhD}}, p_{\mathrm{DPE}}),
\end{equation}
\begin{equation}
    N_{\mathrm{S1, PE}} \sim N_{\mathrm{PhD}} + N_{\mathrm{DPE}}.
\end{equation}
However, the $\mathrm{S1}$ is not always equal to the number of PEs due to reconstruction bias.
This is modeled as
\begin{equation}
    \mathrm{S1} / N_{\mathrm{S1, PE}} - 1 \sim \mathrm{Norm}(\delta_{\mathrm{S1}}, \Delta_{\mathrm{S1}}).
\end{equation}
Here, both $\delta_{\mathrm{S1}}$, $\Delta_{\mathrm{S1}}$ are functions of $N_{\mathrm{PhD}}$ obtained from the XENONnT MC.
In the end, the relative spatial correction is applied back to the $\mathrm{S1}$ signal, yielding
\begin{equation}
    \mathrm{cS1} = \mathrm{S1} \cdot \mathrm{S1corr}(\x',\y',\z),
\end{equation}
where $\x', \y'$ are the event positions as reconstructed in data, which are smeared by the S2-size-dependent resolution of the position reconstruction.

To get the $\mathrm{S2}$ signal from $N_{\mathrm{e}}$, the first step is to simulate the electron loss during the drift.
The survival probability due to attachment to electronegative impurities in LXe is given by
\begin{equation}
    p_{\mathrm{loss}}(\z) = \mathrm{e}^{-\z/(\tau v)},
\end{equation}
where $\tau$ is the electron drift survival time (``electron lifetime'') and $v$ is the drift velocity of electrons in LXe.
Inside the FV the electric field is uniform enough to approximate $v$ as a constant.
Electron losses due to drift field effects close to the wall are accounted for via a spatially dependent charge-insensitive-volume (CIV) probability function $p_{\mathrm{CIV}}(\radius,\z)$ from field simulations of XENONnT~\cite{aprile_field_cage_2023}.
The number of surviving electrons is then given by
\begin{equation}
    N_{\mathrm{surv}} \sim \mathrm{Binom}\left(N_{\mathrm{e}},\: p_{\mathrm{loss}}(\z) \cdot p_{\mathrm{CIV}}(\radius,\z)\right).
\end{equation}
At the liquid-gas interface, a fraction of electrons are extracted,
\begin{equation}
    N_{\mathrm{extr}} \sim \mathrm{Binom}(N_{\mathrm{surv}}, \epsilon(\x,\y)).
\end{equation}
The extraction efficiency is $\x$-$\y$ dependent and can be calculated by
\begin{equation}
    \epsilon(\x, \y) = g_2 \cdot \mathrm{S2corr}^{-1}(\x, \y) / G,
\end{equation}
where $g_2$ is the averaged ionization gain, $\mathrm{S2corr}$ is the relative spatial correction factor, and $G\sim 31\,\mathrm{PE/e^-}$ is the single electron gain.
Assuming the fluctuation of the secondary scintillation process is Poisson-like, the expected number of $\mathrm{S2}$ PEs is
\begin{equation}
    N_{\mathrm{S2, PE}} \sim \mathrm{Norm}(N_{\mathrm{extr}}G, \sqrt{N_{\mathrm{extr}}G}).
\end{equation}
Similar to $\mathrm{S1}$, the $\mathrm{S2}$ reconstruction bias is modeled by
\begin{equation}
    \mathrm{S2} / N_{\mathrm{S2, PE}} - 1 \sim \mathrm{Norm}(\delta_{\mathrm{S2}}, \Delta_{\mathrm{S2}}).
\end{equation}
Finally, the correction is applied to the $\mathrm{S2}$,
\begin{equation}
    \mathrm{cS2} = \mathrm{S2} \cdot \mathrm{S2corr}(\x',\y') \cdot \mathrm{e}^{\z/\tau' v}.
\end{equation}
Here, $\tau'$ is the electron lifetime extracted from data, which is used to define the correction. Note that $\tau'$ can be slightly different from the true value $\tau$ due to small variations over time or non-uniformities along the drift path of electrons. The ratio $\tau'/\tau$ is one of the parameters to fit in the band fitting.
In principle this effect can be absorbed into an additional smearing of S2, however in the fitting framework it is still considered separately from others.

For both S1 and S2 signals, data quality selection efficiencies are applied which depend on the respective signal sizes.
For S1 signals, we additionally apply the S1 reconstruction efficiency which is a function of the number of detected photons $N_{\mathrm{PhD}}$.

For NR multi-scatter events, the simulation is slightly altered.
Here the total number of detected photons is defined as the sum of the detected photons for each single energy deposition $i$,
\begin{equation}
    N_{\mathrm{PhD}} = \sum_iN_{\mathrm{PhD}}^{(i)}.
\end{equation}
Similarly, we sum over the number of S2 PEs of only those scatters that contribute to the S2,
\begin{equation}
    N_{\mathrm{S2,PE}} = \sum_{i_{\mathrm{S2}}} N_{\mathrm{S2,PE}}^{(i_{\mathrm{S2}})}.
\end{equation}
Reconstruction biases, signal corrections, and selection efficiencies are correspondingly applied to the merged signals.

\section{Details on the Calibration Fit Results}

The parameters in the ER and NR band fits are summarized in \cref{tab:bbf_param_summary}, showing both the applied prior and the marginal posterior values.
Parameters where prior values are given without uncertainties are fixed in the fit.
The prior for $\langle N_{\mathrm{ex}}/N_{\mathrm{i}}\rangle$ is a uniform distribution in the stated range.
For the posteriors, the stated values and uncertainties correspond to the median and the central 68\% of the marginal posterior distributions.

\begin{table}[b]
  \centering
  \caption{Prior and marginal posterior distributions of each parameter in the ER and NR emission models.}
  \begin{adjustbox}{max width=\textwidth}
    \renewcommand{\arraystretch}{1.2}
    \begin{tabular}{lccr}
    \hline\hline
    Parameter & Prior & Marginal posterior & Unit \\
    \hline
    $W$       & $13.7\pm0.2$ & $13.7^{+0.2}_{-0.2}$ & eV \\
    $f$       & 0.059                & 0.059                & - \\
    \hline
    \multicolumn{4}{c}{ER parameters}                                       \\
    \hline
    $\langle N_{\mathrm{ex}}/N_{\mathrm{i}}\rangle$   & $0.06$ - $0.20$ & $0.13^{+0.04}_{-0.04}$ & -\\
    $\gamma$   & free  & $0.13^{+0.03}_{-0.02}$    & -   \\
    $\delta$   & free  & $0.34^{+0.07}_{-0.07}$    & -   \\
    $\omega$   & free  & $57^{+15}_{-12}$          & keV \\
    $q_0$      & free  & $1.32^{+0.17}_{-0.20}$    & keV \\
    $q_1$      & free  & $0.47^{+0.07}_{-0.05}$    & keV \\
    $q_2$      & free  & $0.030^{+0.002}_{-0.002}$ & -   \\
    $q_3$      & free  & $0.47^{+0.40}_{-0.31}$    & keV \\
    \hline
    \multicolumn{4}{c}{NR parameters} \\
    \hline
    $\zeta$     & 0.047                         & 0.047                     & - \\
    $\delta$    & 0.062                         & 0.062                     & - \\
    $\alpha$    & free                          & $0.92^{+0.07}_{-0.06}$    & - \\
    $\gamma$    & free                          & $0.016^{+0.001}_{-0.001}$ & - \\
    $\beta$     & $239^{+56}_{-18}$             & $334^{+40}_{-43}$         & - \\
    $\kappa$    & $0.139^{+0.006}_{-0.005}$     & $0.138^{+0.005}_{-0.006}$ & - \\
    $\eta$      & $3.3^{+10.6}_{-1.4}$          & $10.0^{+6.8}_{-5.9}$      & - \\
    $\lambda$   & $1.14^{+0.90}_{-0.18}$        & $1.40^{+0.61}_{-0.38}$    & - \\
    \hline\hline
    \end{tabular}
  \end{adjustbox}
  \label{tab:bbf_param_summary}
\end{table}

\end{document}